\documentclass[aps,prd,onecolumn,showpacs,showkeys,amsmath,amssymb]{revtex4}
\usepackage{graphicx}
\usepackage{multirow}
\usepackage{dcolumn}
\usepackage{bm}
\begin{document}

%=====================================================================================
%=====================================================================================
\title{The Decay Properties of the $1^{-+}$ Hybrid State}
%=====================================================================================
%=====================================================================================
%
\author{Hua-Xing Chen}
\email{hxchen@rcnp.osaka-u.ac.jp}
\author{Zi-Xing Cai}
\email{zx_cai@pku.edu.cn}
\author{Peng-Zhi Huang}
\email{pzhuang@pku.edu.cn} \affiliation{Department of Physics and
State Key Laboratory of Nuclear Physics and Technology, Peking
University, Beijing 100871, China}
\author{Shi-Lin Zhu}
\email{zhusl@pku.edu.cn}
\affiliation{Department of Physics
and State Key Laboratory of Nuclear Physics and Technology\\
and Center of High Energy Physics, Peking University, Beijing
100871, China }
\begin{abstract}

Within the framework of the QCD sum rules, we consider the
three-point correlation function, work at the limit $q^2\to 0$ and
$m_\pi\to 0$, and pick out the singular term $\sim {1\over q^2}$ to
extract the pionic coupling constants of the $1^{-+}$ hybrid meson.
Then we calculate the decay widths of different modes. The decay
width of the $S$-wave modes $b_1 \pi, f_1\pi$ increases quickly as
the hybrid meson mass and decay momentum increase. But for the low
mass hybrid meson around 1.6 GeV, the $P$-wave decay mode $\rho \pi$
is very important and its width is around 180 MeV, while the widths
of $\eta \pi$ and $\eta^\prime \pi$ are strongly suppressed. We
suggest the experimental search of $\pi_1(1600)$ through the decay
chains at BESIII: $e^+e^- \rightarrow J/\psi (\psi')\to \pi_1
+\gamma$ or $e^+e^- \rightarrow J/\psi (\psi')\to \pi_1 +\rho$ where
the $\pi_1$ state can be reconstructed through the decay modes
$\pi_1\to \rho\pi\to \pi^+\pi^-\pi^0$ or $\pi_1\to f_1(1285)\pi^0$.
It is also interesting to look for $\pi_1$ using the available
BELLE/BABAR data through the process $e^+e^-\to \gamma^\ast\to
\rho\pi_1, b_1\pi_1, \gamma \pi_1$ etc.

\end{abstract}
\maketitle
\pagenumbering{arabic}

%%%%%%%%%%%%%%%%%%%%%%%%%%%%%%%%%%
\section{Introduction}
%%%%%%%%%%%%%%%%%%%%%%%%%%%%%%%%%%

A hybrid meson state contains one valence quark, one valence
anti-quark and one valence
gluon~\cite{Meyer:2010ku,Klempt:2007cp,Asner:2008nq,Zhu:2007wz},
which is in contrast with the $\bar q q$ meson in the conventional
quark model. Especially the hybrid states with $J^{PC} =0^{--},
0^{+-}, 1^{-+}, 2^{+-}, \cdots$ are of particular interests since
these exotic quantum numbers arise from the manifest gluon degree of
freedom and can not be accessed by the $\bar q q$ meson.
Experimental confirmation of the exotic hybrid mesons and glueballs
will be a direct test of quantum chromodynamics in the low energy
sector. In the past three decades there have been a lot of
experimental and theoretical investigations of the hybrid states.
However, their nature remains elusive.

Up to now there are three candidates with $J^{PC} = 1^{-+}$:
$\pi_1(1400)$, $\pi_1(1600)$ and $\pi_1(2000)$. According to PDG,
their masses and widths are ($1376\pm17$, $300\pm40$) MeV,
($1653^{\Large +18}_{\Large -15}$, $225^{\Large +45}_{\Large -28}$)
MeV and ($2014 \pm 20 \pm 16$, $230 \pm 21 \pm 73$)
MeV~\cite{Amsler:2008zzb}. Many collaborations reported evidence of
the $1^{-+}$ states such as the CLAS Collaboration, the Crystal
Barrel Collaboration, the COMPASS Collaboration, the E862
Collaboration etc.. For example, the $\pi_1(1400)$ state was
observed in the reactions $\pi^- p \rightarrow \eta \pi^0
n$~\cite{Adams:2006sa}, $\bar p p \rightarrow \pi^0 \pi^0 \eta$,
$\bar p n \rightarrow \pi^- \pi^0 \eta$~\cite{Abele:1999tf}, $\pi^-
p \rightarrow \eta \pi^- p$~\cite{Thompson:1997bs}. The
$\pi_1(1600)$ state was observed in the reactions $\pi^- p
\rightarrow \eta^\prime \pi^- p$~\cite{Ivanov:2001rv}, $\pi^- p
\rightarrow \omega \pi^- \pi^0 p$~\cite{Lu:2004yn} and $\pi^- p
\rightarrow \eta \pi^+ \pi^- \pi^- p$~\cite{Kuhn:2004en}. The
$\pi_1(2015)$ meson was observed in the channels $\pi^- p
\rightarrow \omega \pi^- \pi^0 p$~\cite{Lu:2004yn} and $\pi^- p
\rightarrow \eta \pi^+ \pi^- \pi^- p$~\cite{Kuhn:2004en}. Some other
recent experimental papers include
Refs.~\cite{Dzierba:2005jg,Nozar:2008be,Eugenio:2008zz,Alekseev:2009xt}.
We note that the Ref.~\cite{Alekseev:2009xt} studied the $\pi^-
\pi^- \pi^+$ final state whose data show a significant production of
$\pi_1(1600)$ decaying to $\rho \pi$. We also note that the
existence of $\pi_1(2015)$ is still questionable, and it is not
included in the most recent PDG 2010~\cite{Nakamura:2010zzi}.

There have been many theoretical calculations of the hybrid meson
mass in
literature~\cite{Ebert:2009ub,Kim:2008qh,Ping:2009zza,Kitazoe:1983xx,Dudek:2010wm,Dudek:2009qf,Kisslinger:2009pw}.
The $1^{-+}$ mass extracted from the quenched lattice QCD simulation
ranges from 1.74 GeV~\cite{Hedditch:2005zf} and 1.8
GeV~\cite{Bernard:2003jd} to 2 GeV~\cite{McNeile:1998cp}. Among the
various phenomenological models, the flux tube model is a popular
one. Within this framework the glue degree of freedom is modeled by
a semi-classical color flux tube. The hybrid meson with the quantum
numbers $J^{PC} = 1^{-+}$ was found to be around 1.9
GeV~\cite{Isgur:1984bm,Burns:2006wz}. The mass from constituent
gluon model is around 1.93 GeV~\cite{Iddir:2007dq}. Many authors
studied the mass of the hybrid meson using the QCD sum rule
formalism~\cite{Jin:2002rw,Govaerts:1984bk,Chetyrkin:2000tj,Huang:1998zj,Latorre:1985tg,Yang:2007cc,Reinders:1981ww,Narison:2009vj}.
In our previous papers, we used the tetraquark currents ($qq\bar q
\bar q$) to study the $1^{-+}$ mesons using the method of QCD sum
rule since both the tetraquark and hybrid interpolating currents may
couple to the same states. The extracted mass is around 1.6 GeV and
2 GeV, quite close to the $\pi_1(1600)$ and
$\pi_1(2000)$~\cite{Chen:2008ne,Chen:2008qw}.

To learn more about the hybrid states, it is equally important to
study their decay properties besides their mass spectrum. In the
flux tube model, a hybrid meson decays as the flux tube breaks. The
``$^3P_0$ pair creation'' mechanism is introduced to create a
quark-antiquark pair with $S = 1$, $L = 1$ and total angular
momentum $J = 0$. With this assumption, the lowest-lying hybrid
state with $J^{PC}=1^{-+}$ cannot decay into two ground states, such
as $\pi \pi$, $\pi \eta$ and $\pi \rho$, etc in the original flux
tube model~\cite{Isgur:1985vy}. Later some authors modified this
phenomenological model further with the introduction of a new decay
vertex in order to study the hybrid meson decay
process~\cite{Godfrey:2002rp,Cawlfield:2005ra}. With this
modification, the contribution of the $P$-wave decay mode $\rho \pi$
is not negligible. There are also other theoretical approaches on
the decay, photo- and electro-production of the $1^{-+}$ hybrid
meson~\cite{Latorre:1985tg,McNeile:2006bz,DeViron:1985xn,Zhu:1998sv,Zhu:1999wg,Zhang:2002id,Page:1998gz,Close:1994pr,Afanasev:1997fp,Szczepaniak:2001qz}.

In literature, the three-point correlation function was invoked to
discuss the decay width of the modes $\rho\pi, \eta\pi$
etc~\cite{DeViron:1985xn,Latorre:1985tg}, where the sum rules were
derived at the symmetric point $p^2=(p^\prime+q)^2=q^2=-Q^2 >0$.
Later, the light cone QCD sum rules was employed to calculate the
decays of the hybrid mesons by one of the present
authors~\cite{Zhu:1998sv,Zhu:1999wg}.

In order to perform a systematical study of the $1^{-+}$ hybrid
meson we use a different formalism within the framework of QCD sum
rule in this paper, which was first developed in
Ref.~\cite{Reinders:1982hd}. After calculating the three-point
correlation functions using the method of operator product
expansion, we ignore the small pion mass term $m_\pi^2$ in the
denominator and pick out the divergent term ${1\over q^2}$ in the
limit $q^2\to 0$ where $q$ is the pion momentum. After comparing the
singular term ${1\over q^2}$ of the three-point correlation function
both at the phenomenological and quark-gluon level, we make Borel
transformation of the variables $p^2, {p'}^2=(p+q)^2$, and extract
the coupling constants (such as $g_{\rho \pi}$). Then we calculate
the decay widths of both the isovector hybrid states $\pi_1$ of
$I^GJ^{PC} = 1^-1^{-+}$ and the isoscalar ones $\sigma_1$ of
$I^GJ^{PC} = 0^+1^{-+}$. The non-vanishing decay modes include
$\pi_1 \rightarrow \rho \pi,~\eta \pi,~\eta^\prime \pi,~b_1 \pi,~f_1
\pi$, and $\sigma_1 \rightarrow \eta \eta^\prime,~a_1 \pi,~f_1
\eta$, etc.

Our paper is separated into several sections according to the
different decay modes. In Sec.~\ref{sec:rhopi}, we study the decay
mode $\pi_1 \rightarrow \rho \pi$. In Sec.~\ref{sec:etapi}, we
study the decay modes $\pi_1 \rightarrow \eta \pi$ and
$\eta^\prime \pi$. In Se.~\ref{sec:b1piDE}, we study the decay
mode $\pi_1 \rightarrow b_1 \pi$ using the derivative current. In
Sec~\ref{sec:f1pi}, we study the decay mode $\pi_1 \rightarrow f_1
\pi$. In Sec~\ref{sec:isoscalar}, we extend the same formalism to
study the isoscalar hybrid state with $I^G J^{PC} = 0^+ 1^{-+}$.
Sec.~\ref{sec:summary} is the summary. In
Appendix~\ref{sec:b1piT}, we study the decay mode $\pi_1
\rightarrow b_1 \pi$, using the tensor current for $b_1$.

%%%%%%%%%%%%%%%%%%%%%%%%%%%%%%%%%%%%%%%%%%%%%%%%%%%%%%%%%%%%%%%%%%%%%%%%%
\section{The Decay Mode $\pi_1 \rightarrow \rho \pi$}\label{sec:rhopi}
%%%%%%%%%%%%%%%%%%%%%%%%%%%%%%%%%%%%%%%%%%%%%%%%%%%%%%%%%%%%%%%%%%%%%%%%%

\subsection{Three-Point Correlation Function}

For the past decades QCD sum rule has proven to be a very powerful
and successful non-perturbative
method~\cite{Shifman:1978bx,Reinders:1984sr}. In the usual QCD sum
rule, we consider the two-point correlation function in order to
extract the hadron masses. In this paper we consider the
three-point correlation functions in order to study the decay
properties of hadrons:
\begin{eqnarray}
T^{A \rightarrow BC}(p, p^\prime, q) = \int d^4x d^4y e^{ip^\prime
x} e ^{i q y} \langle0|{\mathbb T} \eta^{B}(x) \eta^C(y)
\eta^{A\dagger}(0) |0\rangle \, .
\end{eqnarray}

The hybrid state $\pi_1$ has several decay modes, such as $\rho
\pi$, $f_1(1285) \pi$, $b_1(1235) \pi$ etc.. So we need calculate
several three-point correlation functions. The procedures are more
or less the same. In this section we study the decay mode $\pi_1
\rightarrow \rho \pi$. First we show the three currents used in
our calculation. The hybrid current with $J^{PC}=1^{-+}$ is:
\begin{eqnarray}
\eta^{\pi_1}_{\mu} = \bar q^a \gamma^\nu {\lambda_{ab}^n\over2}
g_s G^n_{\mu\nu} q^b\, ,
\end{eqnarray}
where the summation is taken over repeated indices ($\mu$, $\nu,
\cdots$ for Lorentz indices, and $a, b, \cdots$ for color
indices). The hybrid state $\pi_1$ is an isovector state. We study
the neutral one, whose quark configurations are $(\bar u u - \bar
d d) g$:
\begin{eqnarray}
\eta_{\mu} \equiv \eta^{\pi_1^0}_{\mu} = {1\over\sqrt2}( \bar u^a
\gamma_\nu u^b - \bar d^a \gamma_\nu d^b ) { \lambda_{ab}^n \over 2
} g_s G^n_{\mu\nu} \, ,
\end{eqnarray}
and it couples to $\pi_1(1600)$ through
\begin{eqnarray}
\langle0|\eta_{\mu}|\pi_1(p,\lambda)\rangle = \sqrt 2 f_{\pi_1}
m_{\pi_1}^3 \epsilon^{\lambda}_\mu \, .
\end{eqnarray}
For the $\rho$ meson, we use the vector current
\begin{eqnarray}
j_\mu^{\rho^+} = \bar d \gamma_\mu u \, , \,j_\mu^{\rho^0} =
{1\over\sqrt2}(\bar u \gamma_\mu u - \bar d \gamma_\mu d) \, , \,
j_\mu^{\rho^-} = \bar u \gamma_\mu d \, ,
\end{eqnarray}
and it couples to the vector meson $\rho$ through
\begin{eqnarray}
\langle0|j^\rho_\mu|\rho(p,\lambda)\rangle = m_\rho f_\rho
\epsilon_\mu^{\lambda} \, .
\end{eqnarray}
For the pseudoscalar meson $\pi$, we use
\begin{eqnarray}
j_5^{\pi^+} = \bar d \gamma_5 u \, , \,j_5^{\pi^0} =
{1\over\sqrt2}(\bar u \gamma_5 u - \bar d \gamma_5 d) \, , \,
j_5^{\pi^-} = \bar u \gamma_5 d \, ,
\end{eqnarray}
and it couples to $\pi$ through
\begin{eqnarray}
\langle0|j_5^\pi|\pi(p)\rangle = f_\pi^\prime = {2i \langle \bar q q
\rangle \over f_\pi} \, .
\end{eqnarray}

Having defined all these interpolating currents and their
couplings to the physical hadrons, we can write down the
three-point correlation function at the phenomenological side:
\begin{eqnarray}\label{eq:rhopiPH}
T^{\rho\pi({\rm PH})}_{\mu\nu}(p, p^\prime, q) &=& \int d^4x d^4y
e^{ip^\prime x} e ^{i q y} \langle0|{\mathbb T} j^{\rho^-}_{\nu}(x)
j^{\pi^+}(y) \eta^\dagger_\mu(0) |0\rangle
\\ \nonumber &=&  g_{\rho \pi} \epsilon_{\mu^\prime\nu^\prime\alpha\beta} {q^\alpha p^{\prime\beta}}
(g_{\mu\mu^\prime} - {{p_\mu} p_{\mu^\prime} \over m_{\pi_1}^2})
(g_{\nu\nu^\prime} - {{p^\prime_\nu} p^\prime_{\nu^\prime} \over
m_\rho^2}) { \sqrt 2  f_{\pi_1} m_{\pi_1}^3 f_\rho m_\rho
f_\pi^\prime \over (m_{\pi_1}^2 - p^2-i\epsilon) (m_\rho^2 -
p^{\prime2}-i\epsilon)(m_\pi^2 - q^2-i\epsilon)}
\\ \nonumber &=&  g_{\rho \pi} \epsilon_{\mu\nu\alpha\beta} {q^\alpha p^{\prime\beta}}  { \sqrt 2  f_{\pi_1}
m_{\pi_1}^3 f_\rho m_\rho f_\pi^\prime \over (m_{\pi_1}^2 -
p^2-i\epsilon) (m_\rho^2 - p^{\prime2}-i\epsilon)(m_\pi^2 -
q^2-i\epsilon)} \, ,
\end{eqnarray}
where the momenta of $\pi_1$, $\rho$ and $\pi$ are $p_\mu$,
$p^\prime_\mu$ and $q_\mu$, respectively. This is for the decay of
$\pi_1^0 \rightarrow \rho^- \pi^+$, and we can obtain the same
result for $\pi_1^0 \rightarrow \rho^+ \pi^-$. The coupling constant
$g_{\rho \pi}$ is defined as
\begin{eqnarray}
{\cal L}=g_{\rho \pi}\epsilon_{\mu\nu\alpha\beta}\pi_1^{0\mu}
\partial^\alpha\pi^+ \partial^\beta \rho^{-\nu} +\cdots \, .
\end{eqnarray}

To simplify our calculation, we work at the massless pion pole. Such
a formalism was first applied to calculate the pion nucleon coupling
constants very successfully decades
ago~\cite{Reinders:1982hd,Kim:2000ty,Kim:1999tc,Reinders:1984sr}.
Later the same formalism was employed to discuss various pionic
couplings ~\cite{Brito:2004tv}. We first take the small $\pi$ mass
in the denominator to be zero. Then we work at the limit $q^2
\rightarrow 0$. $T^{\rho\pi(\rm PH)}_{\mu\nu}$ is divergent at
$q^2=0$ up to the order of $(q^2)^{-1}$. Therefore, we will also
choose such divergent terms at the QCD side in the next subsection.

\subsection{Operator Product Expansion}

In the previous subsection we have obtained the expression of the
three-point correlation function at the phenomenological side. In
this subsection we will calculate this at the QCD side using the
method of operator product expansion (OPE). To do this, first we
insert the three currents into the three-point correlation
function. After doing the Wick contractions we obtain:
\begin{eqnarray}
&& \langle0|{\mathbb T} j^{\rho^-}_{\nu}(x) j^{\pi^+}(y)
\eta^\dagger_{\mu}(0) |0\rangle
\\ \nonumber &=& {1\over2\sqrt 2} {\rm Tr}\Big ( i S_d^{cd}(x-y) \gamma_5 i
S_u^{da}(y) \gamma_{\mu^\prime} i S_u^{bc}(-x) \gamma_\nu \Big )
\lambda^n_{ab} g_s G_{\mu\mu^\prime}^n(0)
\\ \nonumber && - {1\over2\sqrt 2} {\rm Tr}\Big ( i S_d^{ca}(x) \gamma_{\mu^\prime} i
S_d^{bd}(-y) \gamma_5 i S_u^{dc}(y-x) \gamma_\nu \Big )
\lambda^n_{ab} g_s G_{\mu\mu^\prime}^n(0) \, .
\end{eqnarray}
Inside this expression, $S_u$ and $S_d$ are the quark propagators
for $up$ and $down$ quark, respectively. In the presence of quark
and gluon condensates their expressions are:
%
%%%%%%%%%%%%%%%%%%%%%%%%%%%%%%%%%%%%%%%%%%%%%%%%%%%%%%%%%%%%%%%%%%%%%%%%%%%%%%
\begin{eqnarray}
i S^{ab}(x) & \equiv & \langle 0 | T [ q^a(x) q^b(0) ] | 0 \rangle
\\ \nonumber &=& { i \delta^{ab} \over 2 \pi^2 x^4 } \hat{x} + {i \over
32\pi^2} { \lambda^n_{ab} \over 2 } g_s G^n_{\mu\nu} {1 \over x^2}
(\sigma^{\mu\nu} \hat{x} + \hat{x} \sigma^{\mu\nu}) - { \delta^{ab}
\over 12 } \langle \bar q q \rangle + { \delta^{ab} x^2 \over 192 }
\langle g_c \bar q \sigma G q \rangle \, ,
\end{eqnarray}
%%%%%%%%%%%%%%%%%%%%%%%%%%%%%%%%%%%%%%%%%%%%%%%%%%%%%%%%%%%%%%%%%%%%%%%%%%%%%%
%
where the masses of $up$ and $down$ quarks have been neglected
already.

The OPE calculation is largely simplified when we work at the pion
pole and only choose the terms divergent at the $q^2 \rightarrow
0$ limit:
\begin{eqnarray}\label{eq:rhopiOPE}
T^{\rho\pi({\rm OPE})}_{\mu\nu}(p, p^\prime, q) &=&
{\epsilon_{\mu\nu\alpha\beta} {q^\alpha p^{\prime\beta}} \over q^2
}\Big ( {\langle g_s \bar q \sigma G q \rangle \over 6\sqrt2 } \big(
{3 \over p^2} + { 1 \over p^{\prime 2} } \big )  - {\langle \bar q q
\rangle \langle g_s^2 GG \rangle \over 18 \sqrt2 } \big(  { 1 \over
p^4} + { 1 \over p^{\prime 4}} \big ) \Big ) \, .
\end{eqnarray}
We note here that the tri-gluon condensate $\langle g_s^3 f GGG
\rangle$ as well as the $\alpha_s$ correction vanishes in this
case.

\subsection{Numerical Analysis}

In our numerical analysis, we use the following values for various
condensates and $m_s$ at 1 GeV and $\alpha_s$ at 1.7 GeV
~\cite{Yang:1993bp,Narison:2002pw,Gimenez:2005nt,Jamin:2002ev,Ioffe:2002be,Ovchinnikov:1988gk}:
%
%%%%%%%%%%%%%%%%%%%%%%%%%%%%%%%%%%%%%%%%%%%%%%%%%%%%%%%%%%%%%%%%%%%%%%%%%%%%%%
\begin{eqnarray}
\nonumber &&\langle\bar qq \rangle=-(0.240 \mbox{ GeV})^3\, ,
\\
\nonumber &&\langle\bar ss\rangle=-(0.8\pm 0.1)\times(0.240 \mbox{
GeV})^3\, ,
\\
\nonumber &&\langle g_s^2GG\rangle =(0.48\pm 0.14) \mbox{ GeV}^4\, ,
\\
\nonumber &&\langle g_s^3GGG\rangle = 0.045 \mbox{ GeV}^6\, ,
\\
\label{condensates} && \langle g_s\bar q\sigma G
q\rangle=-M_0^2\times\langle\bar qq\rangle\, ,
\\
\nonumber && M_0^2=(0.8\pm0.2)\mbox{ GeV}^2\, ,
\\
\nonumber &&m_s(1\mbox{ GeV})=125 \pm 20 \mbox{ MeV}\, ,
\\
\nonumber && \alpha_s(1.7\mbox{GeV}) = 0.328 \pm 0.03 \pm 0.025 \, .
\end{eqnarray}
%%%%%%%%%%%%%%%%%%%%%%%%%%%%%%%%%%%%%%%%%%%%%%%%%%%%%%%%%%%%%%%%%%%%%%%%%%%%%%
%
There is a minus sign in the definition of the mixed condensate
$\langle g_s\bar q\sigma G q\rangle$, which is different from that
used in some other QCD sum rule studies. This difference just
comes from the definition of coupling constant
$g_s$~\cite{Yang:1993bp}. We use the following values for the
initial and final
states~\cite{Zhu:1999wg,Reinders:1984sr,Zhu:1998bm}:
\begin{eqnarray}
&& m_{\pi_1} =
1.6 {\rm GeV} \, , f_{\pi_1} = 0.026 {\rm GeV}\, ,  \\
\nonumber && m_\pi = {140 \rm MeV} \, , f_\pi^\prime =
{2i\langle\bar q q\rangle \over f_\pi} = { 2 i (-243{\rm MeV})^3
\over 131 {\rm MeV}} \, , \\ \nonumber && m_\rho = 770 {\rm MeV} \,
, f_\rho = 220 {\rm MeV} \, .
\end{eqnarray}

We can compare the three-point correlation function at the
phenomenological side Eq.~(\ref{eq:rhopiPH}) and at the QCD side
Eq.~(\ref{eq:rhopiOPE}) in order to calculate the coupling
constant $g_{\rho \pi}$. Due to $p = p^\prime + q$, we have
several different choices for the QCD sum rules analysis. Here we
use $p^{\prime}_\mu$ and $q_\mu$,  $p^2$, $p^{\prime 2}$ and $q^2$
instead of $p_\mu$, $p \cdot p^\prime$, $p \cdot q$ and $p^\prime
\cdot q$.

In the QCD sum rule, we use the Borel transformation to suppress the
high order contributions. We use $T$ to denote the Borel Mass $M_B$.
Here we have two types of Borel transformation:
\begin{enumerate}

\item Assume $p^2 = p^{\prime 2}$ and perform the Borel
transformation once, $\mathcal{B}(p^2 = p^{\prime 2} \rightarrow
T^2)$;

\item Perform the Borel transformation twice, $\mathcal{B}(p^2
\rightarrow T_1^2, \, p^{\prime 2} \rightarrow T_2^2)$, and then
assume $T_1 = T_2 = T$.

\end{enumerate}

\begin{figure}[hbt]
\begin{center}
\scalebox{0.6}{\includegraphics{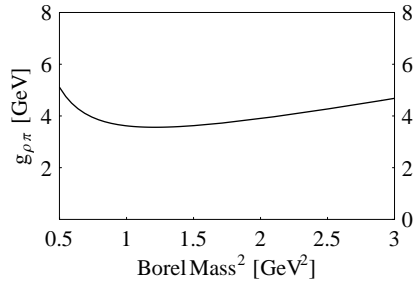}} \caption{The coupling
constant $g_{\rho \pi}$ as a function of $M_B^2$, calculated using
Eq.~({\ref{eq:rhopi})}.} \label{pic:rhopi}
\end{center}
\end{figure}

For the decay mode $\rho \pi$, we employ the first type of Borel
transformation. We compare Eqs.~(\ref{eq:rhopiPH}) and
(\ref{eq:rhopiOPE}), assume $p^2 = p^{\prime 2}$ and perform the
Borel transformation once, $\mathcal{B}(p^2 = p^{\prime 2}
\rightarrow T^2)$ to arrive at
\begin{eqnarray}\label{eq:rhopi}
-g_{\rho \pi} { \sqrt 2  f_{\pi_1} m^3_{\pi_1} f_\rho m_\rho
f_\pi^\prime \over m_\rho^2 - m_{\pi_1}^2 } \Big (
e^{-m_{\pi_1}^2/T^2} - e^{-m_\rho^2/T^2} \Big ) &=& - {2\langle g
\bar q \sigma G q \rangle \over 3\sqrt2 } - {\langle \bar q q
\rangle \langle g_s^2 GG \rangle \over 9 \sqrt2 } {1\over T^2} \,
.
\end{eqnarray}
Using this equation, the coupling constant $g_{\rho \pi}$ can be
calculated as a function of $M_B$, as shown in Fig.~\ref{pic:rhopi}.
The result is around $4$ GeV$^{-1}$. The formula of the decay width
reads
\begin{eqnarray}
\Gamma(\pi_1^0 \rightarrow \rho^+ \pi^- + \rho^- \pi^+) = 2 \times
{g_{\rho \pi}^2 \over 12 \pi } |\vec q_\pi|^3 \, ,
\end{eqnarray}
where $\vec q_\pi$ is the momentum of the final state $\pi$:
\begin{eqnarray} |\vec q_\pi| = { [(m_{\pi_1}^2 - (m_{\rho} +
m_\pi)^2)(m_{\pi_1}^2 - (m_{\rho} - m_\pi)^2)]^{1/2} \over 2
m_{\pi_1} } \, .
\end{eqnarray}
The decay width of $\pi_1 \rightarrow \rho \pi$ is around 180 MeV.

%%%%%%%%%%%%%%%%%%%%%%%%%%%%%%%%%%%%%%%%%%%%%%%%%%%%%%%%%%%%%%%%%%%%%%
\section{The Decay Modes $\pi_1 \rightarrow \eta \pi, \eta^\prime
\pi$}\label{sec:etapi}
%%%%%%%%%%%%%%%%%%%%%%%%%%%%%%%%%%%%%%%%%%%%%%%%%%%%%%%%%%%%%%%%%%%%%%

\subsection{Three-Point Correlation Function}

In this section we study the decay modes $\pi_1 \rightarrow \eta
\pi$ and $\pi_1 \rightarrow \eta^\prime \pi$. The interpolating
current for the $\eta$ meson is
\begin{eqnarray}\label{eq:eta}
j^\eta = \cos \theta_P j^{\eta_8} -  \sin \theta_P j^{\eta_0} \, ,
\end{eqnarray}
where the mixing angle is $\theta_P = -17
^\circ$~\cite{Amsler:1995td}, and it couples to $\eta$ through
\begin{eqnarray}
\langle0| j^\eta |\eta(p)\rangle = \lambda_\eta \, .
\end{eqnarray}
The $SU(3)$ octet and singlet currents in Eq.~(\ref{eq:eta}) are
\begin{eqnarray}
j^{\eta_8} &=& {1\over\sqrt6} ( { \bar u \gamma_5 u +  \bar d
\gamma_5 d - 2 \bar s \gamma_5 s }) \, ,
\\ \nonumber j^{\eta_0} &=& {1\over\sqrt3} ( { \bar u \gamma_5 u +  \bar d
\gamma_5 d + \bar s \gamma_5 s }) \, .
\end{eqnarray}
The interpolating current for the $\eta^\prime$ meson is
\begin{eqnarray}
j^{\eta\prime} = \sin \theta_P j^{\eta_8} +  \cos \theta_P
j^{\eta_0} \, ,
\end{eqnarray}
and it couples to $\eta^\prime$ through
\begin{eqnarray}
\langle0| j^{\eta^\prime}|{\eta^\prime}(p)\rangle =
\lambda_{\eta^\prime} \, .
\end{eqnarray}

The three-point correlation function for the decay mode $\pi_1
\rightarrow \eta \pi$ is:
\begin{eqnarray}\label{eq:etapiPH}
T^{\eta \pi({\rm PH})}_{\mu}(p, p^\prime, q) &=& \int d^4x d^4y
e^{ip^\prime x} e ^{i q y} \langle0|{\mathbb T} j^{\eta}(x) j^\pi(y)
\eta^\dagger_\mu(0) |0\rangle
\\ \nonumber &=& -(g^A_{\eta \pi} q_\alpha + g^B_{\eta \pi} p^\prime_\alpha) (g_{\mu\alpha} - {p_\mu p_\alpha \over m^2_{\pi_1}})
{ \sqrt 2  f_{\pi_1} m_{\pi_1}^3 \lambda_\eta  f_\pi^\prime \over
(m_{\pi_1}^2 - p^2) (m_{\eta}^2 - p^{\prime2})(m_\pi^2 - q^2)} \,
,
\\ \nonumber &=& -(g^A_{\eta \pi} q_\mu + g^B_{\eta \pi} p^\prime_\mu - g^A_{\eta \pi} {p_\mu p \cdot q \over m^2_{\pi_1}}
- g^B_{\eta \pi} {p_\mu p \cdot p^\prime \over m^2_{\pi_1}}) {
\sqrt 2  f_{\pi_1} m_{\pi_1}^3 \lambda_\eta  f_\pi^\prime \over
(m_{\pi_1}^2 - p^2) (m_{\eta}^2 - p^{\prime2})(m_\pi^2 - q^2)} \,
,
\end{eqnarray}
where the coupling constants $g^A_{\eta \pi}$ and $g^B_{\eta \pi}$
are defined in the following Lagrangian
\begin{equation}
\mathcal{L} = i g^A_{\eta \pi} \pi_1^\mu (\partial_\mu \pi) \eta +
i g^B_{\eta \pi} \pi_1^\mu (\partial_\mu \eta) \pi \; .
\end{equation}

In order to calculate the three-point correlation function
$T^{\eta \pi({\rm PH})}_{\mu}$ at the QCD side, once again we work
at the pion pole and only choose the terms divergent at the $q^2
\rightarrow 0$ limit at the quark and gluon level. When $q^2
\rightarrow 0$, we obtain
\begin{eqnarray}\label{eq:etapiOPE}
q^2 T^{\eta \pi({\rm OPE})}_{\mu}(p, p^\prime, q) &\rightarrow&
X(\theta) \times \Big ( q_\mu \big ( p^2 - p^{\prime2} \big ) {
\langle g_s^2 GG\rangle \over 24 \sqrt2 \pi^2} \int_0^\infty
{e}^{\tau_1 p^2 + \tau_2 p^{\prime
2}} {\tau_2 \over 2 (\tau_1 + \tau_2)^2} d\tau_1 d\tau_2 \\
\nonumber && + { \langle g_s^3 f GGG \rangle \over 96 \sqrt2 \pi^2}
\int_0^\infty {e}^{\tau_1 p^2 + \tau_2 p^{\prime 2}} \times \Big (
q_\mu {2 \tau_1 \over (\tau_1 + \tau_2)^2} + p^\prime_\mu { - \tau_2
\over (\tau_1 + \tau_2)^2}
\\ \nonumber && +  q_\mu \big (
p^2 - p^{\prime2} \big ) { -  (\tau_1^2 + \tau_2^2 ) \tau_2 \over
(\tau_1 + \tau_2)^3} + p^\prime_\mu \big ( p^2 - p^{\prime2} \big )
{ -  \tau_2^2  \over (\tau_1 + \tau_2)^2} \Big ) d\tau_1 d\tau_2 \\
\nonumber && - { \langle g_s^3 f GGG \rangle \over 48 \sqrt2 \pi^2}
{p^\mu \over p^2} \Big ) \, ,
\end{eqnarray}
where $X(\theta) = {1\over\sqrt3}\cos \theta_P -
{\sqrt{2\over3}\sin \theta_P}$. In the calculations, we have used
the following equation~\cite{Belyaev:1995ya}
\begin{eqnarray}
\int { e^{i p^\prime x} e^{iqy} d^4 x d^4 y \over (x-y)^{2l} y^{2m}
x^{2n} } = { (-1)^{l+m+n+1} \pi^4 \over 4^{l+m+n-4} l! m! n!}
\int_0^\infty  { e^{\tau_1 p^2 + \tau_2 p^{\prime 2} + \tau_3 q^2}
d\tau_1^l d\tau_2^m d\tau_3^n \over (\tau_1 \tau_2 + \tau_2 \tau_3 +
\tau_3 \tau_1)^{l+m+n-2} } \, .
\end{eqnarray}
We find that only when $i \geqslant j$, the term
\begin{eqnarray} \nonumber
\int_0^\infty { e^{\tau_3 q^2} \tau_3^i  \over (\tau_1 \tau_2 +
\tau_2 \tau_3 + \tau_3 \tau_1)^j }  d\tau_3 \, ,
\end{eqnarray}
is divergent at $q^2 \rightarrow 0$ (it is up to the order of
$(q^2)^{j-i-1}$).

\subsection{Numerical Analysis}

To perform the numerical analysis, we use the following
values~\cite{Zhu:1998bm}
\begin{eqnarray}
&& m_\eta = 0.547 {\rm GeV} \, , \lambda_\eta = 0.23 {\rm GeV}^2 \,
,
\\ \nonumber && m_{\eta^\prime} = 0.958 {\rm GeV} \, , \lambda_{\eta^\prime} = 0.33 {\rm GeV}^2 \, .
\end{eqnarray}
There are two independent Lorentz structures, $q_\mu$ and
$p^\prime_\mu$. We will use both of them to perform the QCD sum rule
analysis.

\subsubsection{Lorentz structure $q_\mu$}

First we choose the Lorentz structure $q_\mu$.
%and
%assume $g^A_{\eta \pi}$ is the dominant one.
If we perform the Borel transformation once, the condensate $\langle
g_s^2 GG \rangle$ vanishes, and only the condensate $\langle g_s^3 f
GGG \rangle$ survives. Consequently, the coupling constants $g_{\eta
\pi}$ is tiny.

If we perform the Borel transformation twice, the condensate
$\langle g_s^2 GG \rangle$ does not vanish, and we obtain
\begin{eqnarray}\label{eq:etapi2}
&& ( g^{\rm A}_{\eta \pi} - g^{\rm B}_{\eta \pi} ) { \sqrt 2
f_{\pi_1} m^3_{\pi_1} \lambda_\eta f_\pi^\prime  } \Big ( {1 \over
2} e^{-m_{\pi_1}^2/T_1^2} e^{-m_\eta^2/T_2^2} + {m^2_\eta \over 2
m^2_{\pi_1} } e^{-m_{\pi_1}^2/T_1^2} e^{-m_\eta^2/T_2^2} \Big ) \\
\nonumber &=& { X(\theta) \langle g_s^2 GG\rangle \over 24 \sqrt2
\pi^2} \Big ( - \big ( {\partial \over
\partial{\tau_1}} - {\partial \over
\partial{\tau_2}} \big ) {\tau_2 \over 2 (\tau_1 + \tau_2)^2} \Big ) \\ \nonumber && +
{ X(\theta) \langle g_s^3 f GGG\rangle \over 96 \sqrt2 \pi^2} \Big (
{2 \tau_1 \over (\tau_1 + \tau_2)^2} + \big ( {\partial \over
\partial{\tau_1}} - {\partial \over
\partial{\tau_2}} \big ) {(\tau_1^2 + \tau_2^2)\tau_2 \over (\tau_1 + \tau_2)^3} \Big ) \Big |_{\tau_1 =
{1/T_1^2},\, \tau_2 = {1/T_2^2}} \, .
\end{eqnarray}
\begin{figure}[hbt]
\begin{center}
\scalebox{0.6}{\includegraphics{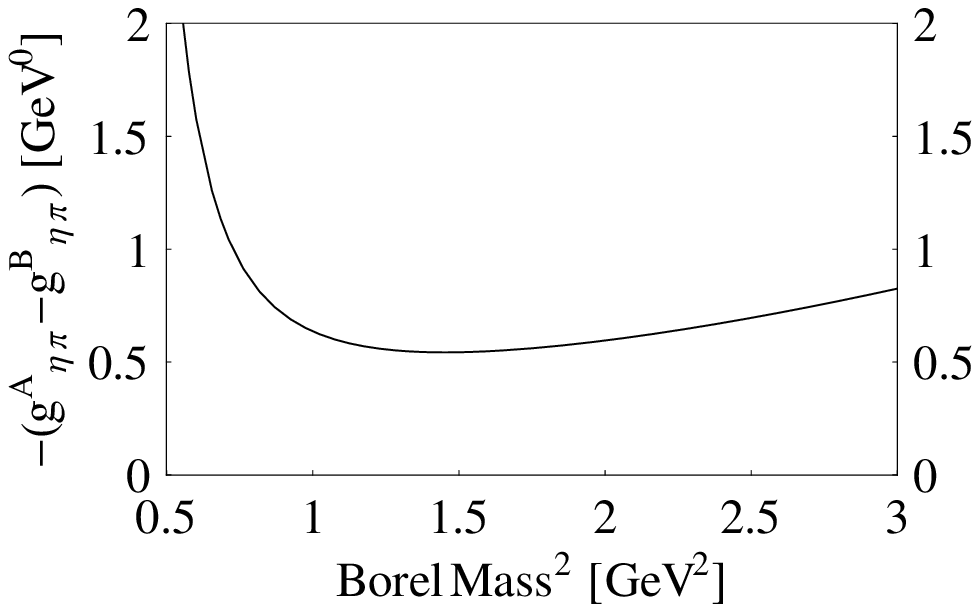}}
\scalebox{0.6}{\includegraphics{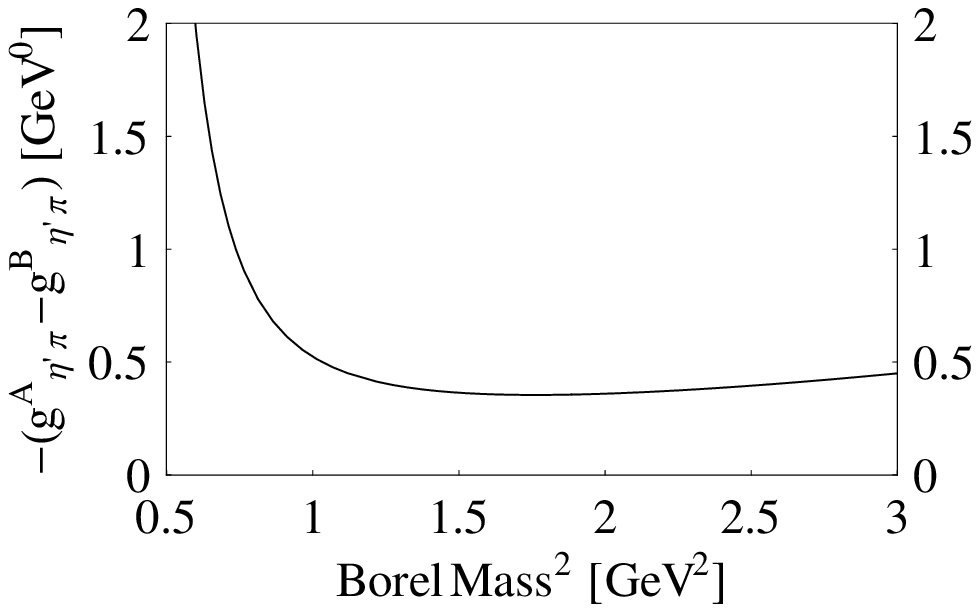}} \caption{The coupling
constant $g_{\eta \pi}$ and $g_{\eta^\prime \pi}$ as functions of
$M_B^2$. The left figure is for $(g^{\rm A}_{\eta \pi} - g^{\rm
B}_{\eta \pi})$, calculated using Eq.~(\ref{eq:etapi2}), and the
right figure is for $(g^{\rm A}_{\eta^\prime \pi} - g^{\rm
B}_{\eta^\prime \pi})$. The Lorentz structure here is $q_\mu$.}
\label{pic:etapi}
\end{center}
\end{figure}
Using Eq.~(\ref{eq:etapi2}) we can calculate $(g^{\rm A}_{\eta \pi}
- g^{\rm B}_{\eta \pi})$. The coupling constant $(g^{\rm
A}_{\eta^\prime \pi} - g^{\rm B}_{\eta^\prime \pi})$ can be obtained
similarly. They are functions of $M_B$, as shown in
Fig.~\ref{pic:etapi}. The L.H.S. is for $(g^{\rm A}_{\eta \pi} -
g^{\rm B}_{\eta \pi})$, and the result is around -0.7; the R.H.S is
for $(g^{\rm A}_{\eta^\prime \pi} - g^{\rm B}_{\eta^\prime \pi})$,
and the result is around -0.4. Using the formula of the decay width
\begin{eqnarray}
\Gamma(\pi_1^0 \rightarrow \eta \pi^0) = {|g^{\rm A}_{\eta \pi} -
g^{\rm B}_{\eta \pi}|^2 \over 24 \pi } {|\vec q_\pi|^3 \over
m_{\pi_1}^2} \, ,
\end{eqnarray}
the decay width of $\pi_1 \rightarrow \eta \pi$ is around 0.9 MeV,
and the one of $\pi_1 \rightarrow \eta^\prime \pi$ is about 0.1 MeV.

\subsubsection{Lorentz structure $p^\prime_\mu$}

Next we move on to the Lorentz structure $p^\prime_\mu$. After
performing the Borel transformation twice, we obtain
\begin{eqnarray}\label{eq:etapi4}
&& - ( g^{\rm A}_{\eta \pi} - g^{\rm B}_{\eta \pi} ) { \sqrt 2
f_{\pi_1} m^3_{\pi_1} \lambda_\eta f_\pi^\prime  } \Big ( {1 \over
2} e^{-m_{\pi_1}^2/T_1^2} e^{-m_\eta^2/T_2^2} - {m^2_\eta \over 2
m^2_{\pi_1} } e^{-m_{\pi_1}^2/T_1^2} e^{-m_\eta^2/T_2^2} \Big ) \\
\nonumber &=& { X(\theta) \langle g_s^3 f GGG\rangle \over 96 \sqrt2
\pi^2} \Big ( {- \tau_2 \over (\tau_1 + \tau_2)^2} + \big (
{\partial \over
\partial{\tau_1}} - {\partial \over
\partial{\tau_2}} \big ) { \tau_2^2 \over (\tau_1 + \tau_2)^2} \Big ) \Big |_{\tau_1 =
{1/T_1^2},\, \tau_2 = {1/T_2^2}} \, .
\end{eqnarray}
However, the condensate $\langle g_s^2 GG \rangle$ vanishes here,
and the obtained result is very small (around -0.05 for $(g^{\rm
A}_{\eta \pi} - g^{\rm B}_{\eta \pi})$).

%%%%%%%%%%%%%%%%%%%%%%%%%%%%%%%%%%%%%%%%%%%%
\section{The Decay Mode $\pi_1 \rightarrow b_1(1235) \pi$}
\label{sec:b1piDE}
%%%%%%%%%%%%%%%%%%%%%%%%%%%%%%%%%%%%%%%%%%%%

\subsection{Three-Point Correlation Function}

In this section we study the decay mode $\pi_1 \rightarrow b_1(1235)
\pi$. For the $b_1(1235)$ meson, we use the following current
\begin{eqnarray}
j_{\mu}^{b_1^+} = (\bar d \overleftrightarrow{D_{\mu}} \gamma_5 u)
\, , j_{\mu}^{b_1^0} = {1\over\sqrt2}(\bar u
\overleftrightarrow{D_{\mu}} \gamma_5 u - \bar d
\overleftrightarrow{D_{\mu}} \gamma_5 d) \, , j_{\mu}^{b_1^-} =
(\bar u \overleftrightarrow{D_{\mu}} \gamma_5 d) \, ,
\end{eqnarray}
and it couples to $b_1(1235)$ through~\cite{Reinders:1984sr}
\begin{eqnarray}
\langle0|j^{b_1}_\mu|b_1(1235)(p,\lambda)\rangle = f_{b_1}
\epsilon^{\lambda}_\mu \, .
\end{eqnarray}
Then we can write down the three-point correlation function at the
phenomenological side:
\begin{eqnarray}\label{eq:b1piDEPH}
T^{b_1 \pi {\rm (PH)}}_{\mu\nu}(p, p^\prime, q) &=& \int d^4x d^4y
e^{ip^\prime x} e ^{i q y} \langle0|{\mathbb T} j^{b_1^-}_{\nu}(x)
j^{\pi^+}(y) \eta^\dagger_\mu(0) |0\rangle
\\ \nonumber &=&
(g_{\mu\mu^\prime}- {p_\mu p_{\mu^\prime} \over
m_{\pi_1}^2})(g_{\nu\nu^\prime}-{p^\prime_{\nu}
p^\prime_{\nu^\prime} \over m_{b_1}^2})  { \sqrt 2 f_{\pi_1}
m_{\pi_1}^3 f_{b_1} f_\pi^\prime \over (m_{\pi_1}^2 - p^2)
(m_{b_1}^2 - p^{\prime2})(m_\pi^2 - q^2)}
\\ \nonumber && \times \Big ( g^{\rm
S}_{b_1 \pi} g_{\mu^\prime \nu^\prime} + g^{\rm D}_1 ( (p \cdot
p^\prime) g_{\mu^\prime \nu^\prime} - p^\prime_{\mu^\prime}
p_{\nu^\prime})  + g^{\rm D}_2 ( (p \cdot q) g_{\mu^\prime
\nu^\prime} - q_{\mu^\prime} p_{\nu^\prime}) \\ \nonumber && +
g^{\rm D}_3 ( (p^\prime \cdot q) g_{\mu^\prime \nu^\prime} -
p^\prime_{\mu^\prime} q_{\nu^\prime})  + g^{\rm D}_4 ( q^2
g_{\mu^\prime \nu^\prime} - q_{\mu^\prime} q_{\nu^\prime})  \Big )
\, .
\end{eqnarray}
This is for the decay mode $\pi_1^0 \rightarrow b_1^- \pi^+$. The
same result holds for $\pi_1^0 \rightarrow b_1^+ \pi^-$. The
coupling constants $g^{\rm S}_{b_1 \pi}$ and $g^{\rm D}_i$
(i=1,2,3,4) are defined through the following effective
Lagrangians:
\begin{eqnarray}
\mathcal{L}_{\rm S} = g^{\rm S}_{b_1 \pi} \vec \pi_{1\alpha} \times
\vec b_{1\beta} \cdot \vec \pi g^{\alpha \beta} \, ,
\end{eqnarray}
\begin{eqnarray}
\mathcal{L}_{\rm D} &=& + g^{\rm D}_1 ( \partial_\alpha \vec
\pi_{1\beta} -
\partial_\beta \vec \pi_{1\alpha} ) \times \partial^\alpha \vec b_1^\beta \cdot \vec \pi
\\ \nonumber && + g^{\rm D}_2 ( \partial_\alpha \vec \pi_{1\beta} -
\partial_\beta \vec \pi_{1\alpha} ) \times \vec b_1^\beta \cdot \partial^\alpha \vec \pi
\\ \nonumber && + g^{\rm D}_3 \vec \pi_{1}^\beta \times ( \partial_\alpha \vec b_{1\beta} - \partial_\beta \vec
b_{1\alpha} ) \cdot \partial^\alpha \vec \pi
\\ \nonumber && + g^{\rm D}_4 (\vec \pi_{1\beta} \times \vec b_{1}^{\beta} \cdot \partial_\alpha \partial^\alpha
\vec \pi - \vec \pi_{1\beta} \times \vec b_{1\alpha} \cdot
\partial^\alpha
\partial^\beta \vec \pi ) \, ,
\end{eqnarray}

Once again, the OPE calculation is largely simplified when we work
at the pion pole and only choose those divergent terms at $q^2
\rightarrow 0$. When $q^2 \rightarrow 0$, we obtain
\begin{eqnarray}\label{eq:b1piDEOPE}
q^2 T^{b_1 \pi \rm (OPE)}_{\mu\nu}(p, p^\prime, q) &\rightarrow&
{\langle g_s^2 GG\rangle \over 48 \sqrt2 \pi^2} \int_0^\infty
{e}^{\tau_1 p^2 + \tau_2 p^{\prime 2}} \times \Big ( - q_\mu q_\nu
{\tau_1 \over (\tau_1 + \tau_2)^3}
\\ \nonumber && + q_\mu q_\nu ( p^2 - p^{\prime 2} )
{ \tau_1 \tau_2 \over (\tau_1 + \tau_2)^3} + q_\mu p^\prime_{\nu} (
p^2 - p^{\prime 2} ) { \tau_2^2 \over (\tau_1 + \tau_2)^3} \Big )
d\tau_1 d\tau_2 \\ \nonumber && + {\langle g_s^3 f GGG\rangle \over
192 \sqrt2 \pi^2} \int_0^\infty {e}^{\tau_1 p^2 + \tau_2 p^{\prime
2}} \times \Big ( g_{\mu \nu} {\tau_1 \over (\tau_1 + \tau_2)^3} +
q_\mu q_\nu { 6 \tau_1^3 + 6 \tau_1^2 \tau_2 + 6 \tau_1 \tau_2^2
\over (\tau_1 + \tau_2)^4} \\ \nonumber && + q_\mu p^\prime_\nu { 6
\tau_1 \tau_2 + 4 \tau_2^2 \over (\tau_1 + \tau_2)^3} + p^\prime_\mu
q_\nu { - 4 \tau_2 \over (\tau_1 + \tau_2)^2} + p^\prime_\mu
p^\prime_\nu { - 2 \tau_2^2 \over (\tau_1 + \tau_2)^3}
\\ \nonumber && + g_{\mu \nu} \big ( p^2 -p^{\prime 2} \big ) { \tau_2
\over (\tau_1 + \tau_2)^2} + q_\mu q_\nu \big ( p^2 -p^{\prime 2}
\big ) { (- 2 \tau_1^2 + 2 \tau_1 \tau_2 - 2 \tau_2^2 ) \tau_1
\tau_2 \over (\tau_1 + \tau_2)^4} \\ \nonumber && + q_\mu
p^\prime_\nu \big ( p^2 -p^{\prime 2} \big ) { (4 \tau_1 - 2 \tau_2
) \tau_2^3 \over (\tau_1 + \tau_2)^4} + p^\prime_\mu q_\nu \big (
p^2 -p^{\prime 2} \big ) { - 2 \tau_1 \tau_2^2 \over (\tau_1 +
\tau_2)^3} \\ \nonumber && + p^\prime_\mu p^\prime_\nu \big ( p^2
-p^{\prime 2} \big ) { - 2 \tau_2^3 \over (\tau_1 + \tau_2)^3} \Big
) d\tau_1 d\tau_2 -  { \langle g_s^3 f GGG \rangle \over 48 \sqrt 2
\pi^2 } { p_\mu p_\nu \over p^2 } \, .
\end{eqnarray}
We note that we do not include the higher order $\alpha_s$
corrections such as $\alpha_s \langle g_s^2 GG \rangle$, etc.

\subsection{Numerical Analysis}

For numerical analysis we use the following values:
\begin{eqnarray}
&& m_{b_1} = 1.235 {\rm GeV} \, , f_{b_1} (2 {\rm GeV}) = 0.18 {\rm
GeV}^3 \, .
\end{eqnarray}
where $f_{b_1}$ is obtained by using Eq.~(A.20) in
Ref.~\cite{Reinders:1984sr} when assuming $m_{b_1} = 1235$ MeV.

In order to solve all the five coupling constants $g^{\rm S}_{b_1
\pi}$ and $g^{\rm D}_i$ (i=1,2,3,4), we need to find five equations.
Comparing Eqs.~(\ref{eq:b1piDEPH}) and (\ref{eq:b1piDEOPE}), we find
there are just five Lorentz structures which can be used to obtain
five equations and solve these coupling constants. They are
$g_{\mu\nu}$, $p^\prime_{\mu} p_{\nu}$, $q_{\mu} p_{\nu}$,
$p^\prime_{\mu} q_{\nu}$, and $q_{\mu} q_{\nu}$. Among them,
$p^\prime_{\mu} q_{\nu}$ and $q_{\mu} q_{\nu}$ lead to the OPEs
which are much larger than those obtained using others.
Consequently, $g^{\rm D}_2$ and $g^{\rm D}_4$ are also much larger
than other coupling constants. Using this fact we can simply our
calculations a lot.

\subsubsection{Lorentz Structure $g_{\mu\nu}$}

First we choose the Lorentz Structure $g_{\mu\nu}$. After performing
the Borel transformation once, we obtain
\begin{eqnarray}\label{eq:b1piS}
- g^{\rm S}_{b_1 \pi} { \sqrt 2 f_{\pi_1} m^3_{\pi_1} f_{b_1}
f_\pi^\prime \over m_{b_1}^2 - m_{\pi_1}^2 } \Big (
e^{-m_{\pi_1}^2/T^2} - e^{-m_{b_1}^2/T^2} \Big ) - g^{\rm D}_1 {
\sqrt 2 f_{\pi_1} m^3_{\pi_1} f_{b_1} f_\pi^\prime \over m_{b_1}^2 -
m_{\pi_1}^2 } \Big ( m_{\pi_1}^2 e^{-m_{\pi_1}^2/T^2} - m_{b_1}^2
e^{-m_{b_1}^2/T^2} \Big ) && \\ \nonumber =  - { \langle g_s^3 f
GGG\rangle \over 384 \sqrt2 \pi^2} T && \, .
\end{eqnarray}
We find that the OPE part only contains the condensate $\langle
g_s^3 f GGG \rangle$, and so the numerical result turns out to be
quite small. Therefore, we will only keep $g_{\mu\nu}$ term and
omit other terms coming from $g_{\mu^\prime \nu^\prime}
(g_{\mu\mu^\prime}- {p_\mu p_{\mu^\prime} \over
m_{\pi_1}^2})(g_{\nu\nu^\prime}-{p^\prime_{\nu}
p^\prime_{\nu^\prime} \over m_{b_1}^2})$, etc. to simply our
calculation. The three-point correlation function
(\ref{eq:b1piDEPH}) becomes
\begin{eqnarray}\label{eq:b1piSim}
T^{b_1 \pi \rm(PH)}_{\mu\nu}(p, p^\prime, q) &=& { \sqrt 2 f_{\pi_1}
m_{\pi_1}^3 f_{b_1} f_\pi^\prime \over (m_{\pi_1}^2 - p^2)
(m_{b_1}^2 - p^{\prime2})(m_\pi^2 - q^2)} \times \Big ( \\ \nonumber
&& g_{\mu\nu} \big ( g^{\rm S}_{b_1 \pi} + g^{\rm D}_1 (p \cdot
p^\prime) + g^{\rm D}_2 (p \cdot q)  + g^{\rm D}_3 (p^\prime \cdot
q)  + g^{\rm D}_4 q^2
\big ) \\
\nonumber && - g^{\rm D}_1 p^\prime_{\mu} p^\prime_{\nu} - g^{\rm
D}_2 q_{\mu} p^\prime_{\nu} - ( g^{\rm D}_1 + g^{\rm D}_3 )
p^\prime_{\mu} q_{\nu} - ( g^{\rm D}_2 + g^{\rm D}_4 ) q_{\mu}
q_{\nu} + \cdots \Big ) \, .
\end{eqnarray}
Using this three-point correlation function, we will show in the
next subsection that $g^{\rm D}_1$ is zero because the OPE with the
Lorentz structure $p^\prime_{\mu} p^\prime_{\nu}$ vanishes. Then
after some calculations, we find that the coupling constant $g^{\rm
S}_{b_1 \pi}$ is around 0.02 GeV, which is much smaller compared
with the coupling constants $g^{\rm D}_2$ and $g^{\rm D}_4$.

\subsubsection{Lorentz Structures $p^\prime_{\mu} p^\prime_{\nu}$ and
$p^\prime_{\mu} q_{\nu}$}

For the Lorentz structures $p^\prime_{\mu} p^\prime_{\nu}$ and
$p^\prime_{\mu} q_{\nu}$, the OPE side vanishes, and so we simply
obtain
\begin{eqnarray}
g^{\rm D}_1 = 0 \, , {\rm and } \, g^{\rm D}_3 = 0 \, .
\end{eqnarray}

\subsubsection{Lorentz Structure $q_\mu p^\prime_\nu$}

\begin{figure}[hbt]
\begin{center}
\scalebox{0.6}{\includegraphics{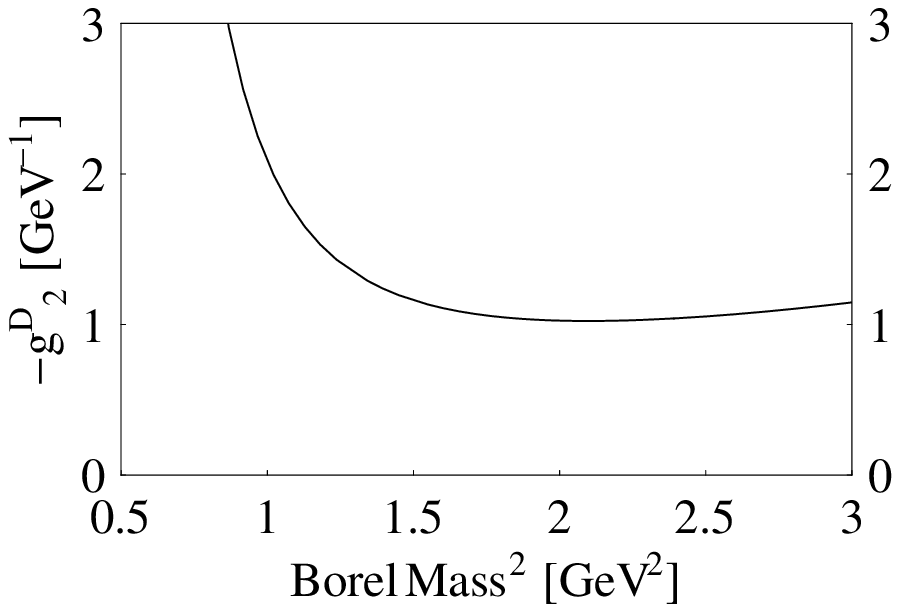}} \caption{The coupling
constant $g^{\rm D}_2$ as a function of $M_B^2$, obtained using
Eq.~(\ref{eq:b1piDED2}). The Lorentz structure is $q_\mu
p^\prime_\nu$.} \label{pic:b1piDED2}
\end{center}
\end{figure}

In this subsection we choose the Lorentz structure $q_\mu
p^\prime_\nu$. After performing the Borel transformation twice, we
obtain
\begin{eqnarray}\label{eq:b1piDED2}
g^{\rm D}_2 { \sqrt 2 f_{\pi_1} m^3_{\pi_1} f_{b_1} f_\pi^\prime
e^{-m_{\pi_1}^2/T_1^2} e^{-m_{b_1}^2/T_2^2} } &=& {\langle g_s^2
GG\rangle \over 48 \sqrt2 \pi^2} \Big ( -\big ( {\partial \over
\partial{\tau_1}} - {\partial \over
\partial{\tau_2}} \big )  {
\tau_2^2 \over (\tau_1 + \tau_2)^3} \Big ) \\ \nonumber &+& {\langle
g_s^3 f GGG\rangle \over 192 \sqrt2 \pi^2}  \Big (  { 6 \tau_1
\tau_2 + 4 \tau_2^2 \over (\tau_1 + \tau_2)^3} -\big ( {\partial
\over
\partial{\tau_1}} - {\partial \over
\partial{\tau_2}} \big )  { (4 \tau_1 - 2 \tau_2 )
\tau_2^3 \over (\tau_1 + \tau_2)^4} \Big ) \Big |_{\tau_1 =
{1/T_1^2},\, \tau_2 = {1/T_2^2}}  \, .
\end{eqnarray}
In Fig.~\ref{pic:b1piDED2} we show the coupling constant $g^{\rm
D}_2$ obtained using this equation, which is around $-1.2$
GeV$^{-1}$.

\subsubsection{Lorentz Structure $q_\mu q_\nu$}

In this subsection we choose the Lorentz structure $q_\mu q_\nu$.
After performing the Borel transformation once, we obtain
\begin{eqnarray}\label{eq:b1piDED241}
( g^{\rm D}_2 + g^{\rm D}_4 ) { \sqrt 2 f_{\pi_1} m^3_{\pi_1}
f_{b_1} f_\pi^\prime \over m_{b_1}^2 - m_{\pi_1}^2 } \Big (
e^{-m_{\pi_1}^2/T^2} - e^{-m_{b_1}^2/T^2} \Big ) &=& {\langle g_s^2
GG\rangle \over 96 \sqrt2 \pi^2} T^2 + { \langle g_s^3 f GGG\rangle
\over 128 \sqrt2 \pi^2} \, .
\end{eqnarray}

We can also perform the Borel transformation twice, and then obtain
\begin{eqnarray}\label{eq:b1piDED242}
&& ( g^{\rm D}_2 + g^{\rm D}_4 ) { \sqrt 2 f_{\pi_1} m^3_{\pi_1}
f_{b_1} f_\pi^\prime
} e^{-m_{\pi_1}^2/T_1^2}  e^{-m_{b_1}^2/T_2^2} \\
\nonumber &=& {\langle g_s^2 GG\rangle \over 48 \sqrt2 \pi^2} \Big (
- {\tau_1 \over (\tau_1 + \tau_2)^3} - \big ( {\partial \over
\partial{\tau_1}} - {\partial \over
\partial{\tau_2}} \big ) { \tau_1 \tau_2 \over (\tau_1 +
\tau_2)^3} \Big ) + {\langle g_s^3 f GGG\rangle \over 192 \sqrt2
\pi^2} \Big ( { 6 \tau_1^3 + 6
\tau_1^2 \tau_2 + 6 \tau_1 \tau_2^2 \over (\tau_1 + \tau_2)^4} \\
\nonumber && - \big ( {\partial \over
\partial{\tau_1}} - {\partial \over
\partial{\tau_2}} \big ) { (- 2 \tau_1^2 + 2 \tau_1 \tau_2 -
2 \tau_2^2 ) \tau_1 \tau_2 \over (\tau_1 + \tau_2)^4} \Big ) \Big
|_{\tau_1 = {1/T_1^2},\, \tau_2 = {1/T_2^2}} \, .
\end{eqnarray}

\begin{figure}[hbt]
\begin{center}
\scalebox{0.6}{\includegraphics{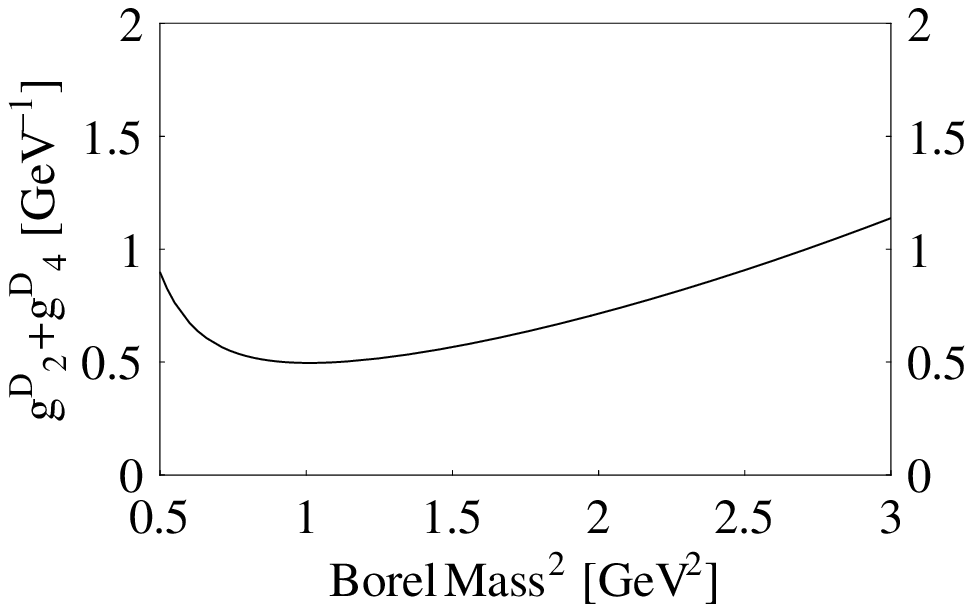}}
\scalebox{0.6}{\includegraphics{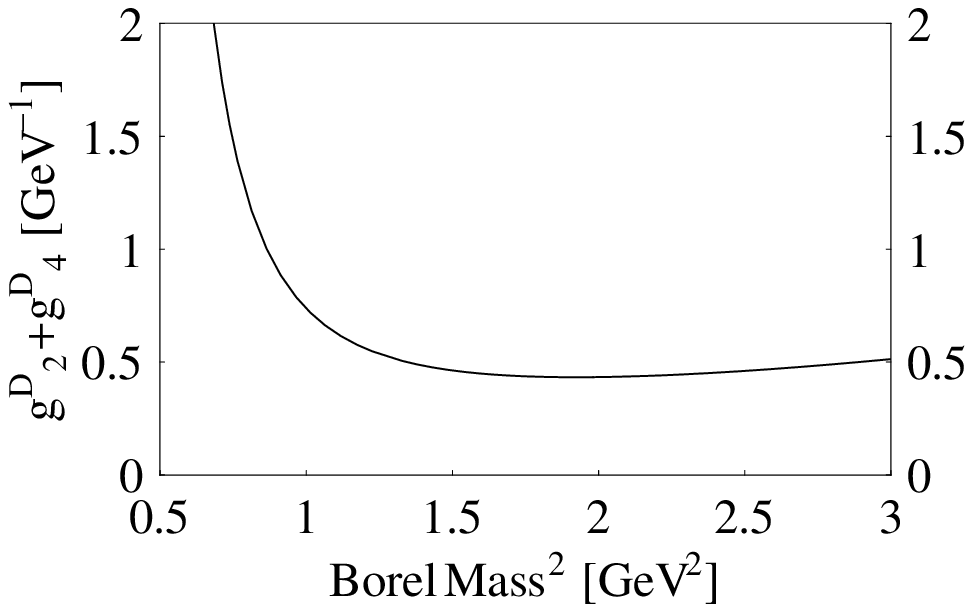}} \caption{The
quantity $g^{\rm D}_2 + g^{\rm D}_4$ as a function of $M_B^2$. The
left and right figures are obtained using Eqs.~(\ref{eq:b1piDED241})
and (\ref{eq:b1piDED242}), respectively. The Lorentz structure is
$q_\mu q_\nu$.} \label{pic:b1piDED24}
\end{center}
\end{figure}

The quantity $g^{\rm D}_2 + g^{\rm D}_4$ is a function of the Borel
mass, as shown in Fig.~\ref{pic:b1piDED24}. The left and right
figures are obtained using Eqs.~(\ref{eq:b1piDED241}) and
(\ref{eq:b1piDED242}), respectively. We obtain that $g^{\rm D}_2 +
g^{\rm D}_4$ is around $0.6$ GeV$^{-1}$, and so the coupling
constant $g^{\rm D}_4$ itself is around $1.8$ GeV$^{-1}$.

\subsection{Decay Width}

The decay width of $\pi_1 \rightarrow b_1 \pi$ reads:
\begin{eqnarray}
\nonumber && \Gamma(\pi_1^0 \rightarrow b_1^+ \pi^- + b_1^- \pi^+)
\\ &=& \label{decay1:b1pi} 2 \times {(g^{\rm S \prime}_{b_1 \pi})^2 \over
24 \pi } {|\vec q_\pi| \over m_{\pi_1}^2}\Big ( 3 + {|\vec q_\pi|^2
\over
m_{b_1}^2} \Big )  \\
&+& \label{decay2:b1pi} 2 \times {(g^{\rm D}_2)^2 \over 24 \pi }
{|\vec q_\pi|^5 \over m_{b_1}^2}
\\
&+& \label{decay3:b1pi} 2 \times {(g^{\rm D}_4)^2 \over 24 \pi }
{|\vec q_\pi|^5 \over m_{\pi}^2 m_{b_1}^2}\Big ( 2 |\vec q_\pi|^2 +
m_{\pi_1}^2 + m_{b_1}^2 + 2 \sqrt{ |\vec q_\pi|^2 + m_{b_1}^2 }
\sqrt{ |\vec q_\pi|^2 + m_{\pi}^2 }  \Big )
\\
&+& \label{decay4:b1pi} 2 \times {g^{\rm S \prime}_{b_1 \pi}g^{\rm
D}_2 \over 12 \pi } {|\vec q_\pi|^3 \over m_{\pi_1} m_{b_1}^2}
\sqrt{ |\vec q_\pi|^2 + m_{b_1}^2 }
\\
&+& \label{decay5:b1pi} 2 \times {g^{\rm S \prime}_{b_1 \pi}g^{\rm
D}_4 \over 12 \pi } {|\vec q_\pi|^3 \over m_{\pi_1}^2 m_{b_1}^2}\Big
( \sqrt{ |\vec q_\pi|^2 + m_{b_1}^2 } \sqrt{ |\vec q_\pi|^2 +
m_{\pi}^2 } + |\vec q_\pi|^2 + m_{b_1}^2 \Big )
\\
&+& \label{decay6:b1pi} 2 \times {g^{\rm D}_2 g^{\rm D}_4 \over 12
\pi } {|\vec q_\pi|^5 \over m_{\pi_1} m_{b_1}^2}\Big ( \sqrt{
|\vec q_\pi|^2 + m_{\pi}^2 } + \sqrt{ |\vec q_\pi|^2 + m_{b_1}^2 }
\Big ) \, ,
\end{eqnarray}
where $g^{\rm S \prime}_{b_1 \pi}$ is defined as
\begin{eqnarray}
g^{\rm S \prime}_{b_1 \pi} = g^{\rm S}_{b_1 \pi} + m_{\pi_1}
\sqrt{m_\pi^2 + |\vec q_\pi|^2} g^{\rm D}_2 + m_\pi^2 g^{\rm D}_4
= - 0.58 {\rm GeV} \, .
\end{eqnarray}
The decay width of $\pi_1 \rightarrow b_1 \pi$ is around $3$ MeV.

%%%%%%%%%%%%%%%%%%%%%%%%%%%%%%%%%%%%%%%%%%%%%%%%%%%%%%%%%
\section{The Decay Mode $\pi_1 \rightarrow f_1(1285)
\pi$}\label{sec:f1pi}
%%%%%%%%%%%%%%%%%%%%%%%%%%%%%%%%%%%%%%%%%%%%%%%%%%%%%%%%%

\subsection{Three-Point Correlation Function}

In this section we study the decay mode $\pi_1 \rightarrow f_1(1285)
\pi$. The isoscalar axial-vector current for $f_1(1285)$ is
\begin{eqnarray}
j^{f_1}_\mu(x)= {1\over\sqrt2}(\bar u \gamma_\mu \gamma_5 u + \bar d
\gamma_\mu \gamma_5 d) \, ,
\end{eqnarray}
and it couples to $f_1(1285)$ through~\cite{Reinders:1984sr}
\begin{eqnarray}
\langle0|j_\mu^{f_1}|f_1(p,\lambda)\rangle = m_{f_1} f_{f_1}
\epsilon^\mu_{f_1} \, .
\end{eqnarray}
Then we can write down the three-point correlation function for the
decay mode $\pi_1 \rightarrow f_1(1285) \pi$:
\begin{eqnarray}\label{eq:f1piPH}
T^{f_1 \pi({\rm PH})}_{\mu\nu}(p, p^\prime, q) &=& \int d^4x d^4y
e^{ip^\prime x} e ^{i q y} \langle0|{\mathbb T} j^{f_1}_{\nu}(x)
j^\pi(y) \eta^\dagger_\mu(0) |0\rangle
\\ \nonumber &=& \big (
g_{\mu \mu^\prime} - {p_\mu p_{\mu^\prime} \over m_{\pi_1}^2} \big )
\big ( g_{\nu \nu^\prime} - {p^\prime_\nu p^\prime_{\nu^\prime}
\over m_{f_1}^2} \big ) { \sqrt 2 f_{\pi_1} m_{\pi_1}^3 f_{f_1}
m_{f_1} f_\pi^\prime \over (m_{\pi_1}^2 - p^2) (m_{f_1}^2 -
p^{\prime2})(m_\pi^2 - q^2)}
\\ \nonumber && \times \Big ( g^{\rm
S}_{f_1 \pi} g_{\mu^\prime \nu^\prime} + g^{\rm D}_1 ( (p \cdot
p^\prime) g_{\mu^\prime \nu^\prime} - p^\prime_{\mu^\prime}
p_{\nu^\prime})   + g^{\rm D}_2 ( (p \cdot q)
g_{\mu^\prime \nu^\prime} - q_{\mu^\prime} p_{\nu^\prime}) \\
\nonumber && + g^{\rm D}_3 ( (p^\prime \cdot q) g_{\mu^\prime
\nu^\prime} - p^\prime_{\mu^\prime} q_{\nu^\prime})  + g^{\rm D}_4 (
q^2 g_{\mu^\prime \nu^\prime} - q_{\mu^\prime} q_{\nu^\prime}) \Big
) \, .
\end{eqnarray}
We note here that the current $j_\mu^{f_1}$ can also couple to the
pseudoscalar meson $\eta$ and $\eta^\prime$ through
\begin{eqnarray}
\langle0|j_\mu^{f_1}|\eta(p), \eta^\prime(p)\rangle = i
f_{\eta,\eta^\prime} p_\mu \, .
\end{eqnarray}
Now the three-point correlation function is:
\begin{eqnarray}\label{eq:f1pi_eta}
T^{f_1 \pi({\rm PH})}_{\mu\nu}(p, p^\prime, q) &=& g^\prime_{\eta
\pi} p^\prime_\nu (g^{\rm C}_{\eta \pi} q_\alpha + g^{\rm D}_{\eta
\pi} p^\prime_\alpha) (g_{\mu\alpha} - {p_\mu p_\alpha \over
m^2_{\pi_1}}) { \sqrt 2 i f_{\pi_1} m_{\pi_1}^3 f_{\eta,
\eta^\prime} f_\pi^\prime \over (m_{\pi_1}^2 - p^2)
(m_{\eta,\eta^\prime}^2 - p^{\prime2})(m_\pi^2 - q^2)} \, .
\end{eqnarray}
Only Eq.~(\ref{eq:f1piPH}) contains the Lorentz structures
$g_{\mu\nu}$, $q_\mu q_\nu$ and $p^\prime_\mu q_\nu$, which can be
used to differentiate them.

At the quark and gluon level, we obtain the following OPE which is
divergent at $q^2 \rightarrow 0$:
\begin{eqnarray}\nonumber\label{eq:f1piOPE}
T^{f_1 \pi({\rm OPE})}_{\mu\nu}(p, p^\prime, q) &=& g_{\mu \nu}
\times \Big ({i \langle g_s \bar q \sigma G q \rangle \over 6\sqrt2
} \big( {3 p \cdot q \over p^2 q^2} + { p^\prime \cdot q \over
p^{\prime 2} q^2 } \big ) + {i \langle \bar q q \rangle \langle
g_s^2 GG \rangle \over 18 \sqrt2 } \big(  - { p \cdot q \over p^4
q^2} + { p \cdot q \over p^2 q^4} - { p^\prime \cdot q \over
p^{\prime 4} q^2 } \\ \nonumber && + { p^\prime \cdot q \over
p^{\prime 2} q^4 } \big ) + {i \langle g_s \bar q \sigma G q \rangle
\langle g_s^2 GG \rangle \over 48 \sqrt2 } \big( - { 1 \over
p^{\prime 4} q^2 } + { 1 \over p^{\prime 2} q^4 } + ({ 1 \over
p^{\prime 6} q^4 } - { 1 \over p^{\prime 4} q^6 })
{8\over3}(p^\prime \cdot q)^2 \big ) \Big ) \\ \nonumber && + q_\mu
q_\nu \times \Big ( - {i \langle g_s \bar q \sigma G q \rangle \over
\sqrt2 }{1\over p^2 q^2} + {i \langle \bar q q \rangle \langle g_s^2
GG \rangle \over 18 \sqrt2 } {1\over p^4 q^2} \Big ) \\ && +
p^\prime_\mu q_\nu \times \Big ( - {i \langle g_s \bar q \sigma G q
\rangle \over 6 \sqrt2 } \big ( {1\over p^{\prime2} q^2} + {3\over
p^2 q^2} \big ) \Big ) \, .
\end{eqnarray}
We note that the $\alpha_s$ correction vanishes in this case.

\subsection{Numerical Analysis}

We use the following values to perform the numerical analysis
~\cite{Reinders:1984sr}:
\begin{eqnarray}
m_{f_1} = 1285 {\rm MeV} \, , f_{f_1} = 170 {\rm MeV} \, .
\end{eqnarray}
where $f_{f_1}$ is obtained using the sum rules Eq.~(4.52) in
Ref.~\cite{Reinders:1984sr} when assuming $m_{f_1} = 1285$ MeV.
Comparing Eqs.~(\ref{eq:b1piDEPH}) and (\ref{eq:b1piDEOPE}), we find
several Lorentz structures, and we can obtain several different QCD
sum rules. They are $g_{\mu\nu}$, $p^\prime_{\mu} p_{\nu}$, $q_{\mu}
p_{\nu}$, $p^\prime_{\mu} q_{\nu}$, and $q_{\mu} q_{\nu}$.

\subsubsection{Lorentz Structure $g_{\mu\nu}$}

First we choose the Lorentz structure $g_{\mu\nu}$. After performing
the Borel transformation once, we obtain
\begin{eqnarray}\label{eq:f1piS}
- g^{\rm S}_{f_1 \pi} { \sqrt 2 f_{\pi_1} m_{\pi_1}^3 f_{f_1}
m_{f_1} f_\pi^\prime \over m_{f_1}^2 - m_{\pi_1}^2 } (
e^{-m_{\pi_1}^2/M^2} - e^{-m_{f_1}^2/M^2} ) - g^{\rm D}_1 { \sqrt 2
f_{\pi_1} m_{\pi_1}^3 f_{f_1} m_{f_1} f_\pi^\prime \over m_{f_1}^2 -
m_{\pi_1}^2 } ( m_{\pi_1} e^{-m_{\pi_1}^2/M^2} - m_{f_1}
e^{-m_{f_1}^2/M^2} )  && \\ \nonumber = - {5 i \langle g_s \bar q
\sigma G q \rangle \langle g_s^2 GG \rangle \over 144 \sqrt2 }
{1\over T^2 } &&  \, .
\end{eqnarray}
Here the dominant condensate ${\langle g_s \bar q \sigma G q
\rangle}$ vanishes after the Borel transformation, and so the
coupling constant $g^{\rm S}_{f_1 \pi}$ is calculated to be around
$0.05$ GeV, which is much smaller than $g^{\rm D}_3$ and $g^{\rm
D}_4$ which we will study in the following subsections. So we are in
the same situation as the previous section. Similarly we omit some
small terms, and the three-point correlation function
(\ref{eq:f1piPH}) is simplified to be
\begin{eqnarray}\label{eq:f1piSim}
T^{f_1 \pi{\rm (PH)}}_{\mu\nu}(p, p^\prime, q) &=& { \sqrt 2
f_{\pi_1} m_{\pi_1}^3 f_{f_1} m_{f_1} f_\pi^\prime \over
(m_{\pi_1}^2 -
p^2) (m_{f_1}^2 - p^{\prime2})(m_\pi^2 - q^2)} \times \Big ( \\
\nonumber && g_{\mu\nu} \big ( g^{\rm S}_{b_1 \pi} + g^{\rm D}_1 (p
\cdot p^\prime) + g^{\rm D}_2 (p \cdot q) + g^{\rm D}_3 (p^\prime
\cdot q)  + g^{\rm D}_4 q^2
\big ) \\
\nonumber && - g^{\rm D}_1 p^\prime_{\mu} p^\prime_{\nu} - g^{\rm
D}_2 q_{\mu} p^\prime_{\nu} - ( g^{\rm D}_1 + g^{\rm D}_3 )
p^\prime_{\mu} q_{\nu} - ( g^{\rm D}_2 + g^{\rm D}_4 ) q_{\mu}
q_{\nu}) + \cdots \Big ) \, .
\end{eqnarray}

\subsubsection{Lorentz Structures $p^\prime_{\mu} p^\prime_{\nu}$ and
$q_{\mu} p^\prime_{\nu}$}

For the Lorentz structures $p^\prime_{\mu} p^\prime_{\nu}$ and
$q_{\mu} p^\prime_{\nu}$, the OPE side vanishes, and so we simply
obtain
\begin{eqnarray}
g^{\rm D}_1 = 0 \, , {\rm and } \, g^{\rm D}_2 = 0 \, .
\end{eqnarray}

\subsubsection{Lorentz Structures $p^\prime_\mu q_\nu$}

\begin{figure}[hbt]
\begin{center}
\scalebox{0.6}{\includegraphics{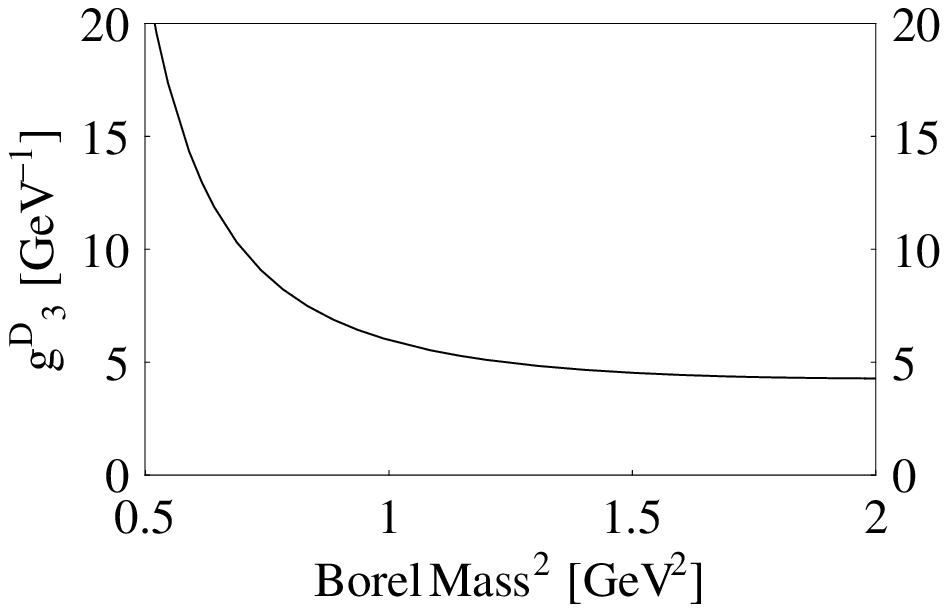}} \caption{The coupling
constant $g^{\rm D}_3$ as a function of $M_B^2$, obtained using
Eq.~(\ref{eq:f1piD3}). The Lorentz structure is $p^\prime_\mu
q_\nu$.} \label{pic:f1piD3}
\end{center}
\end{figure}

In this subsection we choose the Lorentz structure $p^\prime_\mu
q_\nu$. After performing the Borel transformation once, we obtain
\begin{eqnarray}\label{eq:f1piD3}
g^{\rm D}_3 { \sqrt 2 f_{\pi_1} m^3_{\pi_1} f_{f_1} m_{f_1}
f_\pi^\prime \over m_{f_1}^2 - m_{\pi_1}^2 } ( e^{-m_{\pi_1}^2/M^2}
- e^{-m_{f_1}^2/M^2} ) = {2 i \langle g_s \bar q \sigma G q \rangle
\over 3 \sqrt2 } \, .
\end{eqnarray}
The coupling constant $g^{\rm D}_3$ is a function of $M_B$, as shown
in Fig.~\ref{pic:f1piD3}. The result is around $5$ GeV$^{-1}$.

\subsubsection{Lorentz Structures $q_\mu q_\nu$}

\begin{figure}[hbt]
\begin{center}
\scalebox{0.6}{\includegraphics{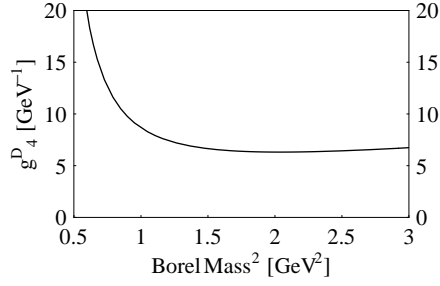}} \caption{The coupling
constant $g^{\rm D}_4$ as a function of $M_B^2$, obtained using
Eq.~(\ref{eq:f1piD4}). The Lorentz structure is $q_\mu q_\nu$.}
\label{pic:f1piD4}
\end{center}
\end{figure}

In this subsection we choose the Lorentz structure $q_\mu q_\nu$.
After performing the Borel transformation once, we obtain
\begin{eqnarray}\label{eq:f1piD4}
g^{\rm D}_4 { \sqrt 2 f_{\pi_1} m^3_{\pi_1} f_{f_1} m_{f_1}
f_\pi^\prime \over m_{f_1}^2 - m_{\pi_1}^2 } ( e^{-m_{\pi_1}^2/M^2}
- e^{-m_{f_1}^2/M^2} ) = {i \langle g_s \bar q \sigma G q \rangle
\over \sqrt2 } + {i \langle \bar q q \rangle \langle g_s^2 GG
\rangle \over 18 \sqrt2} {1\over T^2}\, .
\end{eqnarray}
The coupling constant $g^{\rm D}_4$ is a function of $M_B$, as shown
in Fig.~\ref{pic:f1piD4}. The result is around $7$ GeV$^{-1}$.

\subsection{Decay Width}

To calculate the decay width of $\pi_1 \rightarrow f_1 \pi$, we
similarly introduce
\begin{eqnarray}
g^{\rm S \prime}_{f_1 \pi} = g^{\rm S}_{f_1 \pi} + \big (
\sqrt{m_{f_1}^2 + |\vec q_\pi|^2} \sqrt{m_\pi^2 + |\vec q_\pi|^2} +
|\vec q_\pi|^2 \big ) g^{\rm D}_2 + m_\pi^2 g^{\rm D}_4 = 2.41 {\rm
GeV} \, .
\end{eqnarray}
The decay width reads
\begin{eqnarray}
\nonumber && \Gamma(\pi_1^0 \rightarrow f_1 \pi^0)
\\ &=& \label{decay1:f1pi} {(g^{\rm S \prime}_{f_1 \pi})^2 \over
24 \pi } {|\vec q_\pi| \over m_{\pi_1}^2}\Big ( 3 + {|\vec q_\pi|^2
\over m_{f_1}^2} \Big )  \\
&+& \label{decay2:f1pi} {(g^{\rm D}_3)^2 \over 24 \pi } {|\vec
q_\pi|^5 \over m_{\pi}^2 m_{f_1}^2}\Big ( 2 |\vec q_\pi|^2 +
m_{\pi_1}^2 + m_{f_1}^2 + 2 \sqrt{ |\vec q_\pi|^2 + m_{f_1}^2 }
\sqrt{ |\vec q_\pi|^2 + m_{\pi}^2 }  \Big )
\\
&+& \label{decay3:f1pi} {(g^{\rm D}_4)^2 \over 24 \pi } {|\vec
q_\pi|^5 \over m_{\pi}^2 m_{f_1}^2}\Big ( 2 |\vec q_\pi|^2 +
m_{\pi_1}^2 + m_{f_1}^2 + 2 \sqrt{ |\vec q_\pi|^2 + m_{f_1}^2 }
\sqrt{ |\vec q_\pi|^2 + m_{\pi}^2 }  \Big )
\\
&-& \label{decay4:f1pi} {g^{\rm S \prime}_{f_1 \pi}g^{\rm D}_3
\over 12 \pi } {|\vec q_\pi|^3 \over m_{\pi_1}^2 m_{f_1}^2}\Big (
\sqrt{ |\vec q_\pi|^2 + m_{f_1}^2 } \sqrt{ |\vec q_\pi|^2 +
m_{\pi}^2 } + |\vec q_\pi|^2 + m_{f_1}^2 \Big )
\\
&+& \label{decay5:f1pi} {g^{\rm S \prime}_{f_1 \pi}g^{\rm D}_4
\over 12 \pi } {|\vec q_\pi|^3 \over m_{\pi_1}^2 m_{f_1}^2}\Big (
\sqrt{ |\vec q_\pi|^2 + m_{f_1}^2 } \sqrt{ |\vec q_\pi|^2 +
m_{\pi}^2 } + |\vec q_\pi|^2 + m_{f_1}^2 \Big )
\\
&-& \label{decay6:f1pi} {g^{\rm D}_3 g^{\rm D}_4 \over 12 \pi }
{|\vec q_\pi|^5 \over m_{\pi}^2 m_{f_1}^2}\Big ( 2 |\vec q_\pi|^2 +
m_{\pi_1}^2 + m_{f_1}^2 + 2 \sqrt{ |\vec q_\pi|^2 + m_{f_1}^2 }
\sqrt{ |\vec q_\pi|^2 + m_{\pi}^2 }  \Big ) \, .
\end{eqnarray}
Here (\ref{decay1:f1pi}) mainly comes from the $S$-wave decay,
(\ref{decay2:f1pi}) and (\ref{decay3:f1pi}) come from the $D$-wave
decays, and the last three are the S wave and D wave interference
terms. Numerically they are:
\begin{eqnarray}\nonumber
\mbox{(\ref{decay1:f1pi})} = 23.2 {\rm MeV} \, ,
\mbox{(\ref{decay2:f1pi})} = 0.4 {\rm MeV} \, ,
\mbox{(\ref{decay3:f1pi})} = 0.8 {\rm MeV} \, ,
\mbox{(\ref{decay4:f1pi})} = -2.6 {\rm MeV} \, ,
\mbox{(\ref{decay5:f1pi})} = 3.6 {\rm MeV} \, ,
\mbox{(\ref{decay6:f1pi})} = -1.2 {\rm MeV} \, ,
\end{eqnarray}
The decay width of $\pi_1 \rightarrow f_1 \pi$ is around 24 MeV.

Finally we want to note two important facts here. First, all the
three Lorentz structures $g_{\mu \nu}$, $q_\mu q_\nu$ and
$p^\prime_\mu q_\nu$ which lead to non-zero coupling constants are
not contaminated by the other decay mode $\eta \pi$, as shown in
Eq.~(\ref{eq:f1pi_eta}). Secondly, from Eq.~(\ref{eq:f1piOPE}) we
find that all these three different Lorentz structures contain the
condensate $\langle g_s \bar q \sigma G q \rangle$, the dominant
one. However, for the Lorentz structure $g_{\mu \nu}$, it vanishes
after the Borel transformation. So the obtained $g^{\rm S}_{f_1
\pi}$ is much smaller than $g^{\rm D}_3$ and $g^{\rm D}_4$ obtained
using the other two Lorentz structures, where this condensate does
not vanish.

%%%%%%%%%%%%%%%%%%%%%%%%%%%%%%%%%%%%%%%%%%%%%%%%%%%%%%%%%%%%
\section{Isoscalar Hybrid} \label{sec:isoscalar}
%%%%%%%%%%%%%%%%%%%%%%%%%%%%%%%%%%%%%%%%%%%%%%%%%%%%%%%%%%%%

In this section, we turn to study the decay modes of the isoscalar
hybrid state $\sigma_1$ with the quantum numbers $I^G J^{PC} = 0^+
1^{-+}$. Now the $S$-wave decay modes are $a_1(1260) \pi$ and
$f_1(1285) \eta$, etc., while the $P$-wave decay modes are $\eta
\eta^\prime$, etc. The interpolating current for $\sigma_1$ is
\begin{eqnarray}
\psi_{\mu} = {1\over\sqrt2}( \bar u^a \gamma_\nu u^b + \bar d^a
\gamma_\nu d^b ) { \lambda_{ab}^n \over 2 } g_s G^n_{\mu\nu} \, ,
\end{eqnarray}
and it couples to the isoscalar hybrid state $\sigma_1$ through
\begin{eqnarray}
\langle0|\psi_{\mu}|\sigma_1\rangle = \sqrt 2 f_{\sigma_1}
m_{\sigma_1}^3 \epsilon^{\lambda}_\mu \, .
\end{eqnarray}
The QCD sum rule leads to the following relations
\begin{eqnarray}\label{eq:relation}
T_{\rm OPE}^{\sigma_1 \rightarrow a_1 \pi}(p, p^\prime, q) &=&
T_{\rm OPE}^{\pi_1 \rightarrow f_1 \pi}(p, p^\prime, q) \, ,
\\ \nonumber T_{\rm OPE}^{\sigma_1 \rightarrow f_1 \eta}(p, p^\prime, q)
&=& i \big ({1\over\sqrt3}\cos \theta_P - {\sqrt{2\over3}\sin
\theta_P}\big ) \times T_{\rm OPE}^{\pi_1 \rightarrow f_1 \pi}(p,
p^\prime, q) \, ,
\\ \nonumber T_{\rm OPE}^{\sigma_1 \rightarrow \eta
\eta^\prime}(p, p^\prime, q) &=& i \big ({1\over\sqrt3}\sin \theta_P
+ {\sqrt{2\over3}\cos \theta_P}\big ) \times T_{\rm OPE}^{\pi_1
\rightarrow \eta \pi}(p, p^\prime, q) \, .
\end{eqnarray}
From these relations, the decay width of $\sigma_1 \rightarrow
a_1(1260) \pi \, , \, f_1(1285) \eta \, ,$ and $\eta \eta^\prime$
can be easily obtained. For example, with $m_{\sigma_1} = 2.0 $
GeV and $f_{\sigma_1} = 0.013$ GeV, $m_{a_1} = 1.26$ GeV, $f_{a_1}
= 0.17$ GeV, the decay width of $\sigma_1 \rightarrow a_1(1260)
\pi \, , \, f_1(1285) \eta \, ,$ and $\eta \eta^\prime$ are around
770 MeV, 74 MeV and 0.3 MeV, respectively.

%%%%%%%%%%%%%%%%%%%%%%%%%%%%%%%%%%%%%%%%%%%%%%%%%%%%%%%
\section{Summary} \label{sec:summary}
%%%%%%%%%%%%%%%%%%%%%%%%%%%%%%%%%%%%%%%%%%%%%%%%%%%%%%%

We have studied the decay properties of the hybrid states with
$J^{PC} = 1^{-+}$ using the method of QCD sum rule. We have used
the three-point correlation functions to extract the coupling
constants, and then calculated the relevant decay widths. We work
at the pion pole. First we take the $\pi$ mass in the denominator
to be zero. Then we work at the limit $q^2 \rightarrow 0$ and pick
out the divergent terms only. Such a procedure simplifies the
calculation greatly at the cost of throwing away all the finite
pieces.

%=============================================
\begin{table}[tbh]
\begin{center}
\caption{The decay widths of the isovector hybrid state $\pi_1$ of
$I^GJ^{PC} = 1^-1^{-+}$ and isoscalar one $\sigma_1$ of $I^GJ^{PC}
= 0^+1^{-+}$ when the hybrid mass is taken to be 1.6 GeV, 1.8 GeV
and 2.0 GeV respectively.}
\begin{tabular}{c|c|ccc}
\hline \hline \multirow{2}*{Decay Modes} & \multirow{2}*{Decay
Angular Momentum} & \multicolumn{3}{c}{Widths (MeV)}
\\ \cline{3-5} & & $M=1.6$ GeV & $M=1.8$ GeV & $M=2.0$ GeV \\
\hline
$\pi_1 \rightarrow \rho \pi$ & $P$-wave & $180$ MeV & $410$ MeV & $800$ MeV \\
\hline
$\pi_1 \rightarrow \eta \pi$ & $P$-wave & $0.9$ MeV & $2$ MeV & $4$ MeV \\
\hline
$\pi_1 \rightarrow \eta^\prime \pi$ & $P$-wave & $0.1$ MeV & $0.4$ MeV & $0.9$ MeV \\
\hline $\pi_1 \rightarrow b_1 \pi$ & {$S$+$D$-waves} & $2.9$ MeV &
$14$ MeV & $53$ MeV
\\ \hline  $\pi_1 \rightarrow f_1 \pi$ & {$S$+$D$-waves} &
$24$ MeV & $140$ MeV & $410$ MeV
\\ \hline \hline
$\sigma_1 \rightarrow \eta \eta^\prime$ & $P$-wave & $<0.1$ MeV & $0.1$ MeV & $0.3$ MeV \\
\hline $\sigma_1 \rightarrow a_1 \pi$ & $S$+$D$-waves & $60$ MeV & $310$ MeV & $770$ MeV \\
\hline $\sigma_1 \rightarrow f_1 \pi$ & $S$+$D$-waves & -- & -- & $74$ MeV \\
\hline \hline
\end{tabular}
\label{tab:width}
\end{center}
\end{table}
%=============================================

We have calculated the decay widths of both the isovector hybrid
state $\pi_1$ with $I^GJ^{PC} = 1^-1^{-+}$ and isoscalar one
$\sigma_1$ with $I^GJ^{PC} = 0^+1^{-+}$. The present three-point
correlation function formalism works well for the decay processes
$\pi_1 \rightarrow \rho \pi$, $\pi_1 \rightarrow \eta \pi$ and
$\pi_1 \rightarrow \eta^\prime \pi$ where the Lorentz structures
are relatively simple. We have also discussed the modes $b_1 \pi,
f_1\pi$. These two decay modes are complicated by the many Lorentz
structures in the three-point correlation function, possible
mixing between the S-wave and D-wave decay patterns, and possible
contamination from other decay modes such as $\rho\pi$. We
illustrate the variation of the decay width with the hybrid meson
mass in Table~\ref{tab:width}. The S-wave decay width of the $b_1
\pi, f_1\pi$ modes increases very quickly as the hybrid meson mass
and decay momentum increase. But for the low mass hybrid meson
around 1.6 GeV, the $\rho\pi$ mode is one of the dominant decay
modes.

For the $1^{-+}$ state $\pi_1(1600)$, the decay widths of $\pi_1
\rightarrow \rho \pi$, $\pi_1 \rightarrow \eta \pi$ and $\pi_1
\rightarrow \eta^\prime \pi$ are around 180 MeV, 0.9 MeV and 0.1
MeV, respectively. The modes $\eta \pi, \eta'\pi$ are strongly
suppressed compared with the $\rho \pi$ mode. Moreover, the $\rho
\pi$ mode is one of the dominant decay modes of the $\pi_1(1600)$,
which is in strong contrast with predictions from some
phenomenological models.

In this paper we use the hybrid currents to study these hybrid
states. We can also use the tetraquark currents which may also
couple to these $1^{-+}$ states. We have used such currents to study
their masses, and found their possible decay modes through Fierz
transformation~\cite{Chen:2008ne,Chen:2008qw}. We plan to study
their decay properties like their decay widths etc in the future.
This might be useful to know the internal structure of the $1^{-+}$
states.

We suggest the experimental search of $\pi_1(1600)$ through the
decay chains at BESIII: $e^+e^- \rightarrow J/\psi (\psi')\to \pi_1
+\gamma$ or $e^+e^- \rightarrow J/\psi (\psi')\to \pi_1 +\rho$ where
the $\pi_1$ state can be reconstructed through the decay modes
$\pi_1\to \rho\pi\to \pi^+\pi^-\pi^0$ or $\pi_1\to f_1(1285)\pi^0$.
More details of possible experimental search at BESIII can be found
in Ref. \cite{hpz}. It is also interesting to look for $\pi_1$ using
the available BELLE/BABAR data through the process $e^+e^-\to
\gamma^\ast\to \rho\pi_1, b_1\pi_1, \gamma \pi_1$ etc. Hopefully our
result will be helpful to the experimental identification of the
$1^{-+}$ state.

\acknowledgments{This project was supported by the National
Natural Science Foundation of China under Grants 10625521,
10721063 and Ministry of Science and Technology of China
(2009CB825200). }

\appendix

%%%%%%%%%%%%%%%%%%%%%%%%%%%%%%%%%%%%%%%%%%%%%%%%%%
\section{The Decay Mode $\pi_1 \rightarrow b_1(1235) \pi$ With the Tensor
Current}\label{sec:b1piT}
%%%%%%%%%%%%%%%%%%%%%%%%%%%%%%%%%%%%%%%%%%%%%%%%%%

\subsection{Three-Point Correlation Function}

In this appendix, we consider the tensor current for the
$b_1(1235)$ meson and study the decay mode $\pi_1 \rightarrow
b_1(1235) \pi$:
\begin{eqnarray}\label{eq:b1T}
j_{\mu\nu}^{\rm TE^+} = \bar d \sigma_{\mu\nu} u \, ,
\,j_{\mu\nu}^{\rm TE^0} = {1\over\sqrt2}(\bar u \sigma_{\mu\nu} u
- \bar d \sigma_{\mu\nu} d) \, , \, j_{\mu\nu}^{\rm TE^-} = \bar u
\sigma_{\mu\nu} d \, ,
\end{eqnarray}
which couples to $b_1(1235)$ through~\cite{Jansen:2009yh}
\begin{eqnarray}
\langle 0| j_{\mu\nu}^{\rm TE} |b_1(p,\lambda)\rangle = i
f_{b_1}^{\rm TE} \epsilon_{\mu\nu\alpha\beta}
\epsilon^{\alpha}_{\lambda} p^\beta \, .
\end{eqnarray}
The three-point correlation function at the phenomenological side
is:
\begin{eqnarray}\label{eq:b1piTEph1}
T^{\rm TE(PH)}_{\mu\rho\sigma}(p, p^\prime, q) &=& \int d^4x d^4y
e^{ip^\prime x} e ^{i q y} \langle0|{\mathbb T} j^{\rm
TE^-}_{\rho\sigma}(x) j^{\pi^+}(y) \eta^\dagger_\mu(0) |0\rangle
\\ \nonumber &=&
\epsilon_{\rho\sigma\alpha\beta} p^{\prime\beta} (g_{\mu\mu^\prime}-
{p_\mu p_{\mu^\prime} \over
m_{\pi_1}^2})(g_{\alpha\alpha^\prime}-{p^\prime_{\alpha}
p^\prime_{\alpha^\prime} \over m_{b_1}^2}) { \sqrt 2 i f_{\pi_1}
m_{\pi_1}^3 f^{\rm TE}_{b_1} f_\pi^\prime \over (m_{\pi_1}^2 - p^2)
(m_{b_1}^2 - p^{\prime2})(m_\pi^2 - q^2)}
\\ \nonumber && \times \Big ( g^{\rm
S}_{b_1 \pi} g_{\mu^\prime \nu^\prime} + g^{\rm D}_1 ( (p \cdot
p^\prime) g_{\mu^\prime \nu^\prime} - p^\prime_{\mu^\prime}
p_{\nu^\prime}) )  + g^{\rm D}_2 ( (p \cdot q) g_{\mu^\prime
\nu^\prime} - q_{\mu^\prime} p_{\nu^\prime}) \\ \nonumber && +
g^{\rm D}_3 ( (p^\prime \cdot q) g_{\mu^\prime \nu^\prime} -
p^\prime_{\mu^\prime} q_{\nu^\prime})  + g^{\rm D}_4 ( q^2
g_{\mu^\prime \nu^\prime} - q_{\mu^\prime} q_{\nu^\prime}) ) ) \Big
) \, .
\end{eqnarray}
Unfortunately, the tensor current can also couple to the vector
meson $\rho(770)$ through~\cite{Jansen:2009yh}
\begin{eqnarray}
\langle0| j_{\mu\nu}^{\rm TE} |\rho(p,\lambda)\rangle = i
f_{\rho}^{\rm TE} (\epsilon^\lambda_\mu p_\nu - \epsilon^\lambda_\nu
p_\mu) \, .
\end{eqnarray}
When the tensor current couples to $\rho$, the three-point
correlation function is
\begin{eqnarray}\label{eq:b1piTEph2}
T^{\rm TE(PH)}_{\mu\rho\sigma}(p, p^\prime, q) &=& \int d^4x d^4y
e^{ip^\prime x} e ^{i q y} \langle0|{\rm T} j^{\rm
TE^-}_{\rho\sigma}(x) j^{\pi^+}(y) \eta^\dagger_\mu(0) |0\rangle
\\ \nonumber &=&  (g_{\mu\mu^\prime}-
{p_\mu p_{\mu^\prime} \over m_{\pi_1}^2}) { \sqrt 2 i f_{\pi_1}
m_{\pi_1}^3 f^{\rm TE}_{\rho} f_\pi^\prime \over (m_{\pi_1}^2 - p^2)
(m_{\rho}^2 - p^{\prime2})(m_\pi^2 - q^2)}\\
\nonumber && {\times} g^{\rm TE}_{\rho \pi}
\epsilon_{\mu^\prime\nu^\prime\alpha\beta} q^\alpha p^{\prime\beta}
\Big ( (g_{\rho\nu^\prime}-{p^\prime_{\rho} p^\prime_{\nu^\prime}
\over m_{\rho}^2}) p^\prime_{\sigma} -
(g_{\sigma\nu^\prime}-{p^\prime_{\sigma} p^\prime_{\nu^\prime} \over
m_{\rho}^2}) p^\prime_{\rho} \Big ) \, .
\end{eqnarray}
We note that the following relation exists between these Lorentz
structures
\begin{eqnarray}\label{eq:huang}
\epsilon_{\mu \nu \alpha \beta} q^\alpha p^\beta p_\rho +
\epsilon_{\nu \rho \alpha \beta} q^\alpha p^\beta p_\mu +
\epsilon_{\rho \mu \alpha \beta} q^\alpha p^\beta p_\nu = -
\epsilon_{\mu \nu \rho \alpha} q^\alpha p^2 + \epsilon_{\mu \nu \rho
\alpha} p^\alpha p \cdot q \, .
\end{eqnarray}
And we can write Eq.~(\ref{eq:b1piTEph2}) as
\begin{eqnarray}\label{eq:b1piTEph3}
T^{\rm TE(PH)}_{\mu\rho\sigma}(p, p^\prime, q) &=& g^{\rm TE}_{\rho
\pi} \epsilon_{\mu\rho\sigma\alpha} p^\alpha p \cdot q { \sqrt 2 i
f_{\pi_1} m_{\pi_1}^3 f^{\rm T}_{\rho} f_\pi^\prime \over
(m_{\pi_1}^2 - p^2) (m_{\rho}^2 - p^{\prime2})(m_\pi^2 - q^2)} +
\cdots \, .
\end{eqnarray}
The contribution of $\rho \pi$ appears in our analysis at the
leading Lorentz structure $\epsilon_{\mu \rho \sigma \alpha}
p^{\prime \alpha}$, and makes it difficult to single out the
contribution of $b_1 \pi$.

At the quark and gluon level, we obtain the following OPE which is
divergent at $q^2 \rightarrow 0$
\begin{eqnarray}\label{eq:b1piT}
q^2 T^{\rm TE(OPE)}_{\mu\rho\sigma}(p, p^\prime, q) &\rightarrow&
\epsilon_{\rho \sigma \alpha \beta} p^{\prime \alpha} q^\beta q_\mu
{i \langle g_s^2 GG\rangle \over 24 \sqrt2 \pi^2} \int_0^\infty
{e}^{(\tau_1 + \tau_2) p^2} {\tau_2 \over (\tau_1 + \tau_2)^2}
d\tau_1 d\tau_2 \\ \nonumber && + {i \langle g_s^3 f GGG \rangle
\over 192 \sqrt2 \pi^2} \int_0^\infty {e}^{(\tau_1 + \tau_2) p^2}
\Big ( \epsilon_{\mu \rho \sigma \alpha} q^\alpha \big ( { 8 \tau_1
- 6 \tau_2 \over (\tau_1 + \tau_2 )^2 } + p^{\prime 2} { -8 \tau_2^2
\over (\tau_1 + \tau_2 )^2 } \big )
\\ \nonumber && + \epsilon_{\mu \rho \sigma \alpha} p^{\prime\alpha} { -20 \tau_2
\over (\tau_1 + \tau_2 )^2 } + \epsilon_{\rho \sigma \alpha \beta}
p^{\prime\alpha} q^\beta q_\mu { -4 \tau_2 ( \tau_1^2 + \tau_2^2 )
\over (\tau_1 + \tau_2 )^3 }  \\ \nonumber && + \epsilon_{\mu \rho
\sigma \alpha} p^{\prime\alpha} \big ( p^2 - p^{\prime 2} \big ) { 2
\tau_2^2 \over (\tau_1 + \tau_2 )^2 } \Big ) d\tau_1 d\tau_2 - { 3 i
\langle g_s^3 f GGG \rangle \over 32 \sqrt2 \pi^2} { \epsilon_{\mu
\rho \sigma \alpha} p^\alpha \over p^2 }  \, .
\end{eqnarray}

\subsection{Numerical Analysis}

In the numerical analysis, we use the following
value~\cite{Jansen:2009yh}:
\begin{eqnarray}
f_{b_1}^{\rm TE} (2 {\rm GeV}) = 180(20) {\rm MeV} \, .
\end{eqnarray}

\begin{figure}[hbt]
\begin{center}
\scalebox{0.6}{\includegraphics{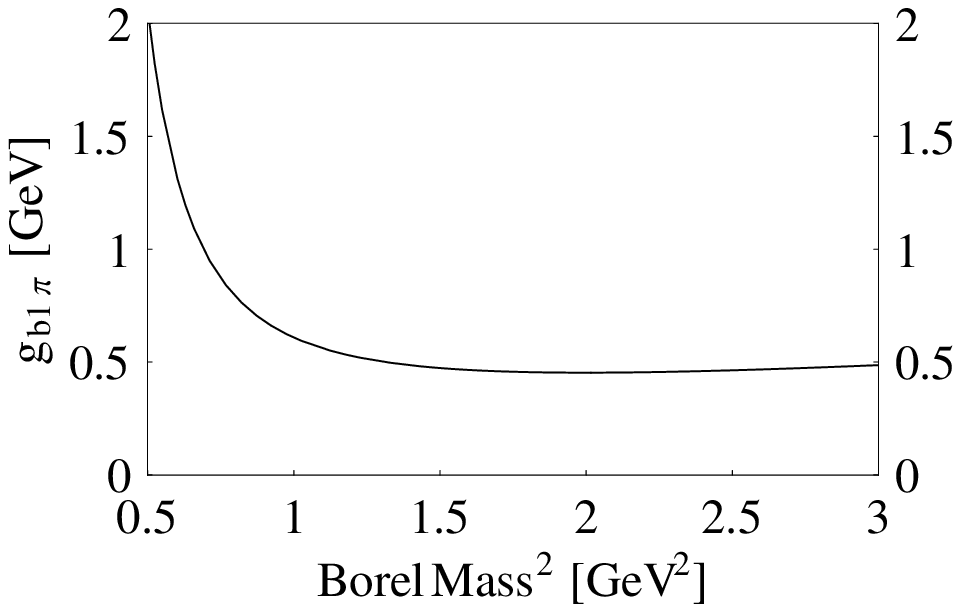}}
\scalebox{0.6}{\includegraphics{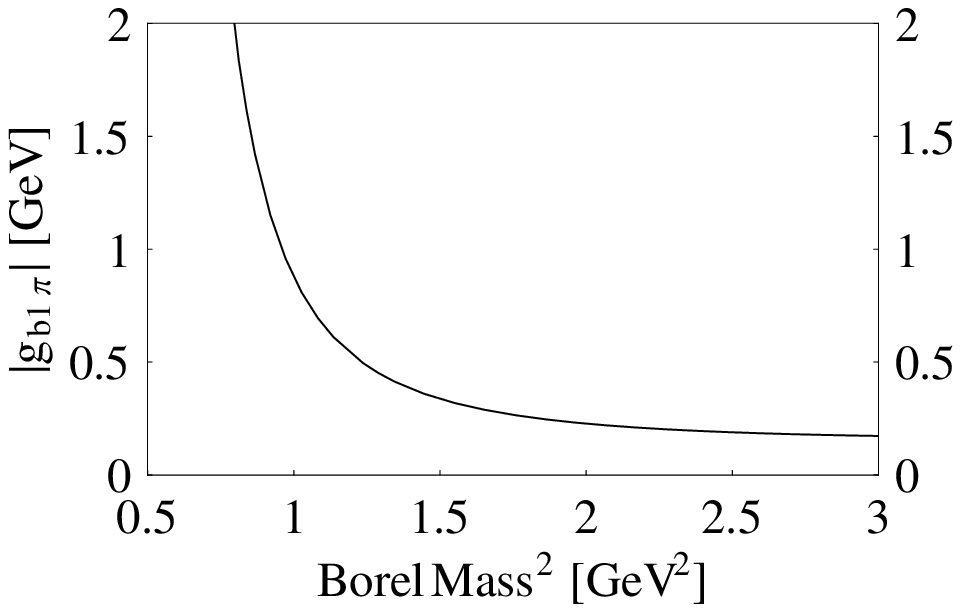}} \caption{The coupling
constant $g^{\rm S}_{b_1 \pi}$ as a function of $M_B^2$. The left
and right figures are obtained using Eqs.~(\ref{eq:b1piTE1}) and
(\ref{eq:b1piTE2}), respectively. The Lorentz structure is
$\epsilon_{\mu \rho \sigma \alpha} p^{\prime \alpha}$.}
\label{pic:b1piTE1}
\end{center}
\end{figure}

The leading Lorentz structure $\epsilon_{\mu \rho \sigma \alpha}
p^{\prime \alpha}$ appears in the $S$-wave decay. Both $\rho \pi$
and $b_1 \pi$ modes contribute to this structure. If we naively
assume that only the $b_1 \pi$ mode contributes, we obtain after
performing the Borel transformation once
\begin{eqnarray}\label{eq:b1piTE1}
- g^{\rm S}_{b_1 \pi} { \sqrt 2 f_{\pi_1} m^3_{\pi_1} f^{\rm
TE}_{b_1} f_\pi^\prime \over m_{b_1}^2 - m_{\pi_1}^2 } \Big (
e^{-m_{\pi_1}^2/T^2} - e^{-m_{b_1}^2/T^2} \Big ) &=&  {7 i \langle
g_s^3 f GGG \rangle \over 48 \sqrt2 \pi^2} \, .
\end{eqnarray}
We can also perform the Borel transformation twice, and then obtain
\begin{eqnarray}\label{eq:b1piTE2}
- g^{\rm S}_{b_1 \pi} { \sqrt 2 f_{\pi_1} m_{\pi_1}^3 f^{\rm
TE}_{b_1} f_\pi^\prime } e^{-m_{\pi_1}^2/T_1^2} e^{-m_{b_1}^2/T_2^2}
&=&  {i \langle g_s^3 f GGG \rangle \over 192 \sqrt2 \pi^2} { -20
\tau_2 \over (\tau_1 + \tau_2 )^2 } \Big |_{\tau_1 = {1/T_1^2},\,
\tau_2 = {1/T_2^2}} \, .
\end{eqnarray}
Using Eqs.~(\ref{eq:b1piTE1}) and (\ref{eq:b1piTE2}) we can
perform the numerical analysis. The extracted $g^{\rm S}_{b_1
\pi}$ is around 0.5 GeV as shown in Fig.~\ref{pic:b1piTE1}, which
is significantly larger than our previous result. We do not use
this value to calculate the decay width since we are unable to
subtract the contamination from the decay mode $\rho \pi$ very
cleanly. One may use the extracted $\rho \pi$ coupling constant in
the previous sections as input and subtract such contribution from
the sum rule derived using the tensor current. However, such an
approach is very crude with large uncertainty.

\end{document}